\documentclass[preprint,pre,aps]{revtex4-2}
\usepackage[dvips]{graphicx}
\usepackage{graphicx}
\usepackage{amsfonts,amsmath,amssymb,amsbsy}
\usepackage{url}
\usepackage{color}
\usepackage{gensymb}
\usepackage{array, makecell}
\usepackage{hyperref}

\usepackage{amssymb}
\usepackage[figuresright]{rotating}

\begin{document}

	\title{Phase behaviour of primitive models of molecular ionic liquids in porous media: effects of  cation shape, ion association and disordered confinement
}
\author{T. Hvozd}
\affiliation{Institute for Condensed Matter Physics of the National
	Academy of Sciences of Ukraine, 1 Svientsitskii St., 79011 Lviv,
	Ukraine}
\author{T. Patsahan\footnote{ the email address of the corresponding author:  taras.patsahan@gmail.com }}
\affiliation{Institute for Condensed Matter Physics of the National
	Academy of Sciences of Ukraine, 1 Svientsitskii St., 79011 Lviv,
	Ukraine\\
	Institute of Applied Mathematics and Fundamental Sciences, Lviv Polytechnic National University, 12 S. Bandera St., 79013 Lviv, Ukraine }
\author{ O. Patsahan}
\affiliation{Institute for Condensed Matter Physics of the National
	Academy of Sciences of Ukraine, 1 Svientsitskii St., 79011 Lviv,
	Ukraine}
\author{ Yu. Kalyuzhnyi}
\affiliation{Institute for Condensed Matter Physics of the National
	Academy of Sciences of Ukraine, 1 Svientsitskii St., 79011 Lviv,
	Ukraine\\
	University of Ljubljana, Faculty of Chemistry and Chemical Technology,  Ve\v{c}na pot 113, SI-1000 Ljubljana, Slovenia	}
\author{M. Holovko}
\affiliation{Institute for Condensed Matter Physics of the National
	Academy of Sciences of Ukraine, 1 Svientsitskii St., 79011 Lviv,
	Ukraine}	

\date{\today}

\begin{abstract}
The phase behaviour of room-temperature ionic liquids (ILs) confined in disordered porous media is studied using a theoretical approach that combines an extension of scaled particle theory, Wertheim’s thermodynamic perturbation theory, and the associative mean spherical approximation. Two models, differing in the shape of the molecular cation, are considered: one with cations formed as charged flexible chains and the other with cations modelled as charged hard spherocylinders. Each model is described by a mixture of dimerized and free ions, while the porous medium is represented as a disordered matrix of hard spheres. We focus on the effects of the molecular cation shape, partial ion association, and disordered confinement on the liquid--vapour-like phase behaviour of the model ILs. In the approximation considered, we find that both the critical temperature $T_{cr}^*$ and critical density $\rho_{cr}^*$ in the model with spherocylinder cations are lower than those in the model with chain cations, and the phase coexistence region is narrower. 
This is the first theoretical attempt to describe an IL model with molecular ions shaped as spherocylinders, particularly in a disordered porous medium.
\end{abstract}

\maketitle

\section{Introduction}\label{sec:1}
 
It is a big pleasure for us to dedicate this article to our good friend and colleague Abdenasser Idrissi on his 60th birthday. Nasser is one of the leading specialists in the study of various molecular liquids, including room-temperature ionic liquids.

Room-temperature ionic liquids (ILs) are compounds with low melting points, composed exclusively of molecular cations and anions. Due to their unique properties such as wide electrochemical window, flexibility in design, low volatility, and non-flammability, ILs are of great importance for many technological applications
\cite{Seddon1997,Fedorov2014}. In these applications,  ILs are often confined in porous materials, for instance,
 as electrolytes in supercapacitors~\cite{Eftekhari2017,Jeanmairet2022,Wu2022,Kondrat2023}. 
The potential applications of nanoconﬁned ILs in supercapacitors, lithium batteries, fuel cells, catalysis, separation, ionogels, carbonisation, and lubrication, among others, are reviewed in~\cite{LeBideau2011,singh2014ionic,Zhang2016}.  
To design any process involving confined ILs, it is necessary to know their thermodynamic properties, including phase equilibria.  
Despite the increasing amount of experimental results, a fundamental understanding of the confinement effect on ILs in nanopores solely through experiments remains incomplete. In particular, from experimental studies, it is difficult to determine the individual effects of the IL properties and the porous matrix characteristics on the phase separation~\cite{Zhang2016,Cheng2024}. In this connection, the development of the theory capable of predicting the effects of disordered confinement on the phase behaviour of ILs still remains a relevant task. 

The vapour--liquid-like phase diagrams of ionic fluids confined in a single-pore geometry were studied numerically by using the density functional theory (DFT)~\cite{Pizio004,Pizio05,Liu2018,Liu2020}  and the field theoretical variational approach~\cite{Loubet2016}. 
In these studies, the ILs were presented either as the restricted primitive model (RPM), i.e., a fluid of oppositely charged hard spheres of the same diameter~\cite{Pizio004,Pizio05,Loubet2016,Liu2018} or as the primitive model of charged hard spheres of different diameters~\cite{Liu2020}. It was shown that the IL below the critical point phase separates into low-density and high-density phases, in analogy to vapour--liquid phase diagram of simple fluids.

In real ILs, the oppositely charged ions are characterised not only by a size disparity, but also by the shape anisotropy and by the location of the charge on the molecular ion. In recent years, a number of primitive models of ILs were proposed \cite{Malvaldi2007,Spohr2009,MartnBetancourt2009,Fedorov2010,Wu2011,Ganzenmller2011,Lindenberg2014,Lindenberg2015,Guzmn2015,SilvestreAlcantara2016,Lu2016,Kalyuzhnyi2018,Yao2023}.
However, as far as we know, the theoretical studies of the phase behaviour in such ionic systems are limited and mainly devoted to the bulk case. 
In particular, the vapour--liquid-like phase diagrams of ILs with chain-like molecular ions were studied theoretically for the bulk case in Ref.~\cite{Guzmn2015,Kalyuzhnyi2018}. 
The model of ILs with chain-like anions of different lengths confined in a slit-like pore of different widths was studied very recently within the framework of the DFT  by incorporating associations between ions with opposite charges \cite{Cheng2024}. In this study, the vapour--liquid phase diagrams depending on both the chain length and the slit width were calculated. 

It should be noted that a single-pore model is oversimplified. 
In a porous medium, in addition to the effects of separate pores, the correlations between the ions confined in different pores become important. Moreover,  disordered porous materials are characterised by specific features such as porosity and pore surface area. 
In \cite{holovko2009highly,chen2009comment,patsahan2011fluids,holovko2012fluids,holovko2012one,holovko2017improvement}, essential progress towards the description of fluids in a porous medium was made within the framework of the scaled particle theory (SPT). The theory allowed one to obtain analytical expressions for thermodynamic functions of hard-sphere (HS) fluids confined in a disordered hard-sphere (HS) matrix. The expressions include three parameters that define the porosity of the matrix. The first one is the geometric porosity $\phi_{0}$ characterising the free volume, which is not occupied by matrix particles. 
The second parameter $\phi$ is the so-called probe particle porosity defined by the chemical potential of a fluid in the limit of infinite dilution. It characterises the adsorption of a fluid in an empty matrix. 
The third parameter $\phi^*$ is defined by the maximum value of the packing fraction of a hard-body fluid in a porous medium. 
Later, the theory was generalised for confined hard-body fluids such as a multicomponent HS mixture  \cite{chen2016scaled} and a mixture of HSs and hard spherocylinders (HSC)~\cite{holovko2017isotropic,Hvozd2018}. The above-mentioned systems of hard-body fluids in an HS matrix can be served as reference systems in the description of more complex systems, for instance, ILs confined in an uncharged porous matrix~\cite{holovko2016vapour,HolPatPat17,holovko2017application,patsahan2018vapor,Hvozd2019,Hvozd2022,Hvozd2024}.

In this work, we extend our previous study of the vapour--liquid-like behaviour in ILs confined in a disordered HS matrix by considering two models which differ in the shape of the molecular cation; namely, we examine the cation modelled as a charged flexible chain and when the cation is modelled as a charged HSC. In both models, the anion is presented as a charged HS. Our goal here is not only to compare the phase diagrams of the confined IL models with flexible and rigid cations but also to elucidate how the porous medium affects the degree of association between cations and anions in these models along the coexistence curve. To this end, we use the theoretical approach developed in~\cite{Hvozd2024}, which combines the SPT theory, Wertheim’s thermodynamic perturbation theory, and the associative mean-spherical approximation (AMSA) and allows one to derive analytical expressions for the thermodynamic functions. Recently, the approach was used to calculate the phase diagrams of the confined IL model with chain-like molecular cations in the limit of full ionic association~\cite{Hvozd2024}. 

The paper is organised as follows. In section 2, we present the models and briefly describe the theoretical formalism. The results are presented and discussed in section~3. We conclude in section 4.

\section{Models and theory }	

\subsection{Models}
We consider two  models of ILs presented in Table~\ref{table1}. In each case, the anion is modelled as a single charged hard sphere (HS) with diameter $\sigma_1$  and charge $z_1=-ez_{-}$ while the molecular cation in model~A and model~B has different shapes. In model~A, the cation is modelled with $m_c$ tangentially bonded HS monomers of the same diameter $\sigma_2$ and with the charge $z_2=ez_{+}$ placed on one of the terminal beads. 
We assume $\sigma_1=\sigma_2=\sigma$ and $z_1=z_2=z=1$. We consider two versions of model~A that correspond to two lengths of the cation chain, version~1 with $m_c=2$ and version~2 with $m_c=3$. 
In model~B, the cation is modelled as a hard spherocylinder (HSC) with length $L$, diameter $\sigma_{sc}=\sigma_1=\sigma_2=\sigma$ and with charge $z_{sc}=ez=e$ placed at the  of one of the hemispherical caps. The length of a HSC, $L$, is defined as the distance between s of its two hemispherical caps. 
This length is chosen to give the spherocylinder cation a size equivalent to the corresponding chain length of the cation in model~A~(Table~\ref{table1}). 
Thus, we consider two versions of model~B corresponding to different HSC lengths $L^*=L/\sigma$: version~1 with $L^*=1$ and version~2 with $L^*=2$. The number densities of ions are $\rho_I=\rho_++\rho_-$, where $\rho_+$ and $\rho_-$ are the number densities of anions and cations, respectively, and $\rho_+=\rho_-$ due to electroneutrality of a whole system. 
Each IL model is explored in the confinement of a disordered porous medium modelled as a matrix of randomly distributed uncharged HS particles of diameter $\sigma_0$.

\begin{table*}[!htb]
	\begin{center}
		\caption{Schematic representation of model~A (a mixture of positively charged chains 
			and negatively charged monomers) and model~B (a mixture of positively charged spherocylinders and negatively charged monomers).}
		\begin{tabular}[t]{|p{35mm}|p{25mm}|p{31mm}|}
			\hline			
			& \makecell{1} & \makecell{2} \\
			\hline
			\makecell{\textbf{model~A} \\ chains + monomers} & \makecell{\;\\ \includegraphics[height=7mm,width=23mm]{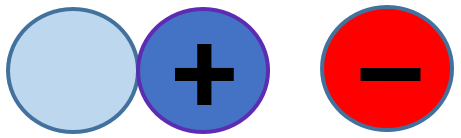}\\$m_c=2$}  &  \makecell{\;\\\includegraphics[height=7mm,width=29mm]{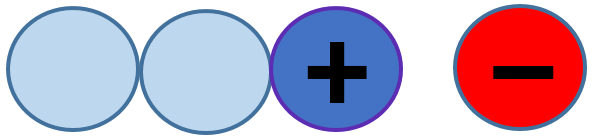}\\$m_c=3$} \\
			\hline
			\makecell{\textbf{model~B} \\ scherocylinders \\ + monomers} & \makecell{\;\\\includegraphics[height=7mm,width=23mm]{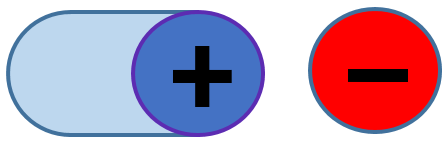}\\$L^*=1$}  &  \makecell{\;\\\includegraphics[height=7mm,width=29mm]{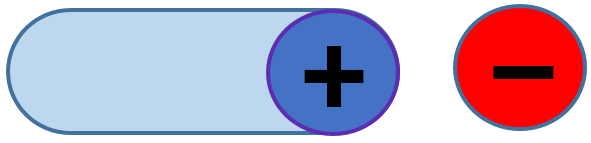}\\$L^*=2$} \\
			\hline
		\end{tabular}\label{table1}
	\end{center}
\end{table*}

We assume that Helmholtz free energy of the IL/matrix system can be presented in the form \cite{holovko2016vapour,holovko2017effects}:
\begin{equation}
\label{F}
\beta F=\beta F^{(id)}+\beta F^{(ref)}+\beta\Delta F,
\end{equation}
where 
$\beta F^{(id)}$ is an ideal-gas contribution,
$\beta F^{(ref)}$ is the free energy of a reference system (RS), 
$\Delta F$ is the contribution connected with the ionic subsystem, $\beta =1/k_{B}T$, $k_B$ is Boltzmann's constant and $T$ is temperature.

It should be emphasised that the RS of the two models are considerably different. 
For model~A, the RS is chosen as a one-component HS system confined in a disordered HS matrix. 
For model~B, a binary mixture of HSs and HSCs confined in a disordered HS matrix is used. 
The ionic subsystems of both models are considered at the same level of approximation. 
Below, we briefly describe the main points of the theoretical approach used to derive analytical 
expressions for the thermodynamic functions of the considered models. 

\subsection{Thermodynamic functions of the reference systems}
\label{subs1}
Here, we present analytical expressions for the thermodynamic functions of the RSs for models~A and B obtained within the framework of the generalised version of the SPT theory \cite{holovko2009highly,chen2009comment,patsahan2011fluids,holovko2012fluids,holovko2012one,holovko2017improvement}.

\subsubsection{A one-component hard-sphere (HS) system confined in a disordered HS matrix} 
\label{par2-2-1}
We start with the RS for model~A presented as a one-component fluid of HSs with diameter $\sigma$ confined in a disordered HS matrix. The number density of the fluid particles is $\rho$. The matrix is characterised by the diameter of HS obstacles $\sigma_0$, their packing fraction $\eta_{0}=\pi\rho_0\sigma_0^3/6$ and three types of porosity: the geometrical porosity $\phi_0=1-\eta_0$, the probe particle porosity $\phi$ determined by the properties of the fluid particle confined in a matrix and the parameter  
$\phi^{\star}$. The expressions for $\phi$ and $\phi^{\star}$ are as follows \cite{holovko2009highly,holovko2017improvement}: 
\begin{align*}
\phi=(1-\eta_0)\exp \bigg[-\frac{3k_{0}\left(1+k_{0}\right)\eta_0}{1-\eta_0}-\frac{9}{2} \frac{k_{0}^2\eta_0^2}{(1-\eta_0)^2}-\frac{k_{0}^3\eta_0}{(1-\eta_0)^3}\left(1+\eta_0+\eta_0^2\right)\bigg],
\end{align*} 
\begin{equation}
\phi^\star=\frac{\phi_0\phi}{\phi_0-\phi}\ln\frac{\phi_0}{\phi},
\label{phi-star}
\end{equation}
where $k_{0}=\sigma/\sigma_0$.

The pressure of this RS can be written as follows:
\begin{align}
\label{pressureCS1}
\beta P^{(\rm{ref})}=\beta P^{\rm{SPT2b3}^\star}+\beta \Delta P^{\rm{CS}},
\end{align}
where $\beta P^{\rm{SPT2b3}^\star}$ is the pressure in the so-called SPT2b3* approximation  \cite{holovko2017improvement}
\begin{align}
\label{pressureSPT2b3star}
\frac{\beta P^{\rm{SPT2b3}^\star}}{\rho}&=
\frac{1}{1-\eta/\phi_0}
+\frac{A}{2}\frac{\eta/\phi_0}{\left(1-\eta/\phi_0\right)^2}
+\frac{2B}{3}\frac{\left(\eta/\phi_0\right)^2}{\left(1-\eta/\phi_0\right)^3}
\nonumber \\
&
+\frac{\phi_0-\phi^\star}{\phi^\star}\frac{\phi_0}{\eta}
\left[\ln(1-\eta/\phi_0)+\frac{\eta/\phi_0}{1-\eta/\phi_0}\right] \nonumber \\
&+\frac{\phi^\star-\phi}{\eta}\left[\ln(1-\eta/\phi^\star)+\frac{\eta/\phi^\star}{1-\eta/\phi^\star}\right]
\end{align}
and $\beta \Delta P^{\rm{CS}}$ is the Carnahan-Starling (CS) correction 
\cite{Carnahan_1969,boublik1974hard}:
\begin{align}
\label{pressCS1-1c}
\frac{\beta \Delta P^{\rm{CS}}}{\rho}=-\frac{\left(\eta/\phi_0\right)^3}{\left(1-\eta/\phi_0\right)^3}.
\end{align}
In the above formulas, the following notations are introduced: $\eta=\pi\rho\sigma^3/6$ is the  volume fraction of the HS fluid particles and
\begin{align*}
A&=6+\frac{3\eta_0 k_0(k_0+4)}{\phi_0}+\frac{9\eta_{0}^{2}k_0^2}{\phi_0^2}, \nonumber \\
B&= \frac{9}{2}\bigg(1+\frac{\eta_{0}k_0}{\phi_0} \bigg)^2.
\end{align*}

Similarly, the chemical potential  can be presented in the form:
\begin{align}
\label{muCSandPL-1}
\beta\mu_i^{(\rm{ref})}=\beta\mu^{\rm{SPT2b3}^\star}+\beta\Delta\mu^{\rm{CS}},
\end{align}
where $\beta\mu^{\rm{SPT2b3}^\star}$ is the chemical potential in the  SPT2b3* approximation \cite{holovko2017improvement}
\begin{align}
\beta\mu^{\rm{SPT2b3}^\star}&=-\ln(1-\eta/\phi_{0})+\frac{\eta/\phi^\star}{1-\eta/\phi_{0}}
+\frac{\eta(\phi^\star-\phi)}{\phi^\star\phi^\star}(1-\eta/\phi^\star)
+A\frac{\eta/\phi_{0}}{1-\eta/\phi_{0}}
\nonumber\\
&+\frac12(A+2B)\frac{(\eta/\phi_{0})^{2}}{(1-\eta/\phi_{0})^{2}} 
+\frac23 B\frac{(\eta/\phi_{0})^{3}}{(1-\eta/\phi_{0})^{3}}
\label{mu_spt2b3-1}
\end{align}
and $\beta\Delta\mu^{CS}$ is the CS correction 
\begin{align}
\label{muCS-1}
\beta\Delta\mu^{CS}=\ln(1-\eta/\phi_0)+\frac{\eta/\phi_0}{1-\eta/\phi_0}-\frac{1}{2}\frac{\left(\eta/\phi_0\right)^2}{\left(1-\eta/\phi_0\right)^2}
-\frac{\left(\eta/\phi_0\right)^3}{\left(1-\eta/\phi_0\right)^3}.
\end{align}
The notations in (\ref{mu_spt2b3-1}) and (\ref{muCS-1}) are the same as in (\ref{pressureSPT2b3star}) and (\ref{pressCS1-1c}).

It is worth noting that the volume fraction of the HS fluid, $\eta$, entering the above equations and the volume fraction of the molecular ions,  $\eta_I$, are connected by the relationship $\eta_I=2\eta/(1+m_c)$, where $m_c$ is the number of monomers in the chain cation.

\subsubsection{A binary mixture of hard spheres (HS) and hard spherocylinders (HSCs) confined in a disordered HS matrix} \label{par2-2-2}

Now we consider the RS for model~B, which is presented as an equimolar mixture of HSs and HSCs confined in a disordered HS matrix. The diameter of HS particles is $\sigma$.  The HSC particles are characterised by their length $L$ and diameter $\sigma$. The volume fraction of the fluid particles is $\eta=\eta_1+\eta_2$, $\eta_{i}=\rho_{i}V_{i}$, where $\rho_{i}$ is the number density of the particles of the $i$-th species, $V_1$ and $V_2$ are volumes of the HS and HSC, respectively:
\begin{equation*}
V_{1}=\frac{\pi}{6}\sigma^3, \qquad 
V_{2}=\frac{\pi}{4}\sigma^2 L+\frac{\pi}{6}\sigma^3.
\end{equation*}

As for a one-component RS, the HS matrix is characterised by the diameter of the HS obstacles $\sigma_0$, the packing fraction $\eta_{0}$ and three types of porosity $\phi_0$, $\phi$ and $\phi^\star$.  In this case, however, the porosity $\phi$ is a function of the probe particle porosity of each species, i.e., $\phi_1$ and $\phi_2$
 \cite{chen2016scaled,Hvozd2018}:
\begin{equation}
\label{smallphip}
\frac{1}{\phi}=\frac{1}{\eta} \sum_{i=1}^{2} \frac{\rho_i V_i}{\phi_i},
\end{equation}
where the expressions for $\phi_1$ and $\phi_2$ are given in Appendix~A, Eqs.~(\ref{probeporosity01})-(\ref{probeporosity02}). Accordingly,  we get $\phi^\star$ (see (\ref{phi-star})) which also depends on $\phi_1$ and $\phi_2$.

The pressure of the system can be presented by Eq.~(\ref{pressureCS1}), where the expression for $\beta P^{\rm{SPT2b3}^\star}$  formally coincides with (\ref{pressureSPT2b3star}), however, in this case,  we have different expressions for $\phi$, $A$ and $B$ (see (\ref{smallphip}) and Appendix~A, Eqs.~(\ref{Ap})-(\ref{b2p})).  

For a mixture confined in a porous medium, the CS correction $\beta \Delta P^{\rm{CS}}$ reads~\cite{Hvozd2018}:
\begin{align}
\label{pressCS1}
 \frac{\beta \Delta P^{\rm{CS}}}{\rho}=-\frac{\left(\eta/\phi_0\right)^3}{\left(1-\eta/\phi_0\right)^3} \Delta_1,
\end{align}
where
\begin{equation}
\Delta_1=\frac{q_m s_m^2}{9 v_m^2}
\label{Delta1}
\end{equation}
with
\begin{align}
\label{vm-sm-qm}
v_m&=\frac{\pi}{6}\sigma^3\left(1+\frac{3L}{4\sigma}\right),\quad
s_m=\pi\sigma^2\left(1+\frac{L}{2\sigma}\right),\nonumber \\
q_m&=\frac{\sigma^2}{4}\left(1+\frac{L^2}{8\sigma^2}+\frac{L}{2\sigma} \right).
\end{align}

It should be noted that we have the partial chemical potentials $\mu_{1}^{(\rm{ref})}$ and $\mu_{2}^{(\rm{ref})}$ corresponding to each species. 
As before, we can write  the chemical potential of the $i$-th species in the form:
\begin{eqnarray}
\label{muCSandPL}
\beta\mu_i^{(\rm{ref})}=\beta\mu_i^{\rm{SPT2b3}^\star}+\beta\Delta\mu_i^{\rm{CS}},
\end{eqnarray}
where the contributions $\beta\mu_i^{\rm{SPT2b3}^\star}$  and $\beta\Delta\mu_i^{\rm{CS}}$ are given by \cite{Hvozd2022}:
\begin{align}
\label{chemSPT2b3star}
&\beta \mu_{i}^{\rm{SPT2b3}^\star}=\beta \mu_{i}^{\rm{SPT2a}}
+\frac{\eta(\phi_0-\phi^\star)}{\phi_0\phi^\star\left(1-\eta/\phi_0\right)}
+\frac{\eta\left(\phi^\star-\phi\right)}{\phi^\star\phi^\star\left(1-\eta/\phi^\star\right)} 
\nonumber\\
&
+\left(\frac{\rho V_i}{\eta}-1\right)
\left[\frac{\phi_0-\phi}{\eta}\ln(1-\eta/\phi_0)+\frac{\phi(\phi_0-\phi^\star)}{\phi_0\phi^\star\left(1-\eta/\phi_0\right)}+\frac{\phi\left(\phi^\star-\phi\right)}{\phi^\star\phi^\star\left(1-\eta/\phi^\star\right)}\right] 
\nonumber \\
&
-\frac{\rho V_i}{\eta}
\left(\frac{\phi}{\phi_i}-1\right)\left[\frac{\phi}{\eta}\ln\left(1-\eta/\phi_0\right)
-\frac{\phi(\phi_0-\phi^\star)}{\phi_0\phi^\star\left(1-\eta/\phi_0\right)}
\right. 
\left.
-\frac{\phi\left(\phi^\star-\phi\right)} {\phi^\star\phi^\star\left(1-\eta/\phi^\star\right)}
+1\right],
\end{align}
\begin{align}
\label{muCS}
\beta\Delta\mu_i^{\rm{CS}}&=-\frac{V_i}{v_m}\frac{\left(\eta/\phi_0\right)^3}{\left(1-\eta/\phi_0\right)^3}\Delta_1+\frac{s_m}{9v_m^3} \big[\left(q_i s_m +2S_i q_m \right)v_m -2V_i q_m s_m\big] \nonumber \\ 
&\times \bigg[\ln(1-\eta/\phi_0)+\frac{\eta/\phi_0}{1-\eta/\phi_0}-\frac{1}{2}\frac{\left(\eta/\phi_0\right)^2}{\left(1-\eta/\phi_0\right)^2}\bigg].
\end{align}
In (\ref{chemSPT2b3star}), $\mu_{i}^{\rm{SPT2a}}$ is the chemical potential of the $i$-th species in the SPT2a approximation \cite{Hvozd2018}. The other notations in (\ref{chemSPT2b3star})-(\ref{muCS}) are the same as for the pressure (see (\ref{pressCS1})-(\ref{vm-sm-qm}) and  Eqs.~ (\ref{probeporosity01})-(\ref{b2p}) in Appendix~A). It should be noted that we will be interested in the sum of the partial chemical potentials $\mu^{(\rm{ref})}=\mu_1^{(\rm{ref})}+\mu_2^{(\rm{ref})}$.  

It is worth noting that we do not take into account the orientation of the HSCs. For the HSC lengths $L^*=1$ and $L^*=2$, the system can be considered isotropic~\cite{frenkel1997}.

\subsection{Ionic subsystem}
\label{subs2}
For model~A, an ionic subsystem consists of an electroneutral mixture of chain cations and monomeric anions immersed in a structureless dielectric medium.   
To describe the thermodynamic properties of the ionic subsystem of model B, we use the approximation in which we approximate a spherocylinder with a charge in the centre of one of the hemisphere caps by a chain of the appropriate length with a charge placed on one of the terminal beads. 
Then, the ionic subsystem of both models is described within the framework of the Wertheim-Orstein-Zernike formalism supplemented by the AMSA \cite{holovko1991effects,Blum95,Protsykevytch1997,KALYUZHNYI1998}.

For each model, the contribution $\beta\Delta F$ to the free energy  (\ref{F})  can be written as a sum of two terms
\[
\beta\Delta f=\frac{\beta\Delta F}{V}=\beta f^{(mal)}+\beta f^{(el)},
\]
where $\beta f^{(mal)}$ is the contribution due to the mass action law (MAL)  and $\beta f^{(el)}$ is the contribution due to electrostatic interactions. For model~A, $\beta f^{(mal)}$ consists of the contribution related to the ionic association between the cations and anions as well as of the contribution related to the formation of the chain cations, i.e. \cite{Bernard96,Jiang02,Kalyuzhnyi2018}, 
\begin{equation}
\label{f-mal}
\beta f^{(mal)}=\beta f^{(ass)}+\beta f^{(ch)}.
\end{equation}
For model~B, $\beta f^{(mal)}$ includes only the contribution $\beta f^{(ass)}$.

The contribution $\beta f^{(ass)}$ is given by
\begin{equation}
	\beta f^{(ass)}=\rho_{I}\left(\ln\alpha-\frac{1}{2}\alpha+\frac{1}{2}\right),
	\label{f-ass}
\end{equation}
where $\rho_{I}=\rho_{+}+\rho_{-}$ is the total number density of ions.  $\alpha$ is the fraction of free anions (or cations) which is determined from the MAL equation 
 \begin{equation}
1-\alpha=\frac{\rho_{I}}{2}\alpha^2 K.
\label{MALEe}
 \end{equation}
In (\ref{MALEe}),  $K=K_{ass}^{(0)}K^{\gamma}$ is the association constant, where the thermodynamic association constant $K_{ass}^{(0)}$ is the infinite-dilute limit of K.  $K^{\gamma}$ is the concentration-dependent part which is determined as the ratio of the activity coefficients of free ions to the activity coefficient of the ion pair. The most common form of the thermodynamic association constant, $K_{ass}^{(0)}=K_{Eb}^{(0)}$, is introduced by Ebeling \cite{Ebeling1968} where $K_{Eb}^{(0)}$ provides an exact second ionic virial coefficient \cite{Ebeling1968}. Here, following \cite{Jiang02,holovko2017effects}, we choose $K_{ass}^{(0)}$ in the form 
proposed by Olaussen and Stell~\cite{Olaussen91} as $K_{ass}^{(0)}\approx 12 K_{Eb}^{(0)}$~\cite{raineri2000phase}. 
 
Using Eq.~(11) from Ref.~\cite{Kalyuzhnyi2018} after some algebra, $K^{\gamma}$ can be presented in the form: 
\begin{equation}
  \label{K_gamma}
  K^{\gamma}=g_{12}^{(hs)}(\sigma_{12}^+)\exp\left(-\frac{1}{T^*}\frac{[\Gamma\sigma(2+\Gamma\sigma)+(\eta_{m_c}^B\sigma^2)^2]}{(1+\Gamma\sigma)^2}\right),
\end{equation}
  where the indices $1$ and $2$ correspond to the anion and the charged monomer of the cation, respectively. $g_{12}^{(hs)}(\sigma_{12}^+)$ is the contact value of the pair distribution function at zero charges,  $m_c$ is the number of monomers in the cation chain, $T^*$  is the dimensionless temperature
  \begin{equation}
 \label{T}
T^*=\frac{k_{B}T\varepsilon\sigma}{\rm{e}^2},
\end{equation} 
$\Gamma$ is  Blum's screening parameter and $\eta_{m_c}^B$ is the parameter describing a shape asymmetry of the cation. If $\eta_{m_c}^B=0$, one arrives at  $K^{\gamma}$  obtained for the RPM in the AMSA \cite{holovko2017effects}.

In (\ref{K_gamma}), the screening parameter $\Gamma$ is the solution of the Wertheim-Orstein-Zernike 
 equation supplemented by the AMSA \cite{bernard1996binding,Bernard2000,Kalyuzhnyi2001,Kalyuzhnyi2018}. Using the results of Ref.~\cite{Kalyuzhnyi2018} we can present the equation for $\Gamma$  as follows:
  \begin{align}	
4\sigma^2\Gamma^2(1+\sigma\Gamma)^3 &=\kappa^2(\alpha+\sigma\Gamma) 
-\displaystyle\frac{\kappa^2\sigma^2\eta^B_{m_c}}{2^{m_c}(1+\sigma\Gamma)^{m_c-1}}\left[F_1^{(m_c)}(\sigma\Gamma,\alpha)\right.
\left.
-\sigma^2\eta^B_{m_c}F_2^{(m_c)}(\sigma\Gamma,\alpha)
\right],	
\label{Gamma}
\end{align}	
where
\begin{equation}
\sigma^2\eta^B_{m_c}=\frac{(1-\Delta^{(m_c)})[2(1+\sigma\Gamma)-(1-\alpha)]f^{(m_c)}(\sigma\Gamma,\alpha)}{D^{(m_c)}(\Delta^{(m_c)},\sigma\Gamma,\alpha)},
\label{eta_B}
\end{equation}	
\begin{equation}
\Delta^{(m_c)}=1-\frac{(m_c+1)}{2}\eta_I, \qquad
\eta_I=\eta_++\eta_-,
\label{Delta}
\end{equation}
 $\kappa=\kappa_D/\sigma$, where $\kappa_D$ is the inverse Debye screening length, $\eta_I$ is the volume fraction of the molecular ions. It should be noted that the functions $F_1^{(m_c)}$, $F_2^{(m_c)}$, $f^{(m_c)}$, $\Delta^{(m_c)}$ in Eqs.~(\ref{Gamma})-(\ref{Delta}) have different forms for different lengths of the cation chain. 
Explicit expressions for these functions for the cations made of two and three monomers are presented below.
We have for model~A  with the cations made of two  monomers:
\begin{align}	
&F_1^{(2)}(\sigma\Gamma,\alpha)=4(1+\sigma\Gamma)-3(1-\alpha) \nonumber \\
&F_2^{(2)}(\sigma\Gamma,\alpha)=4(1+\sigma\Gamma)+6(1+\sigma\Gamma)^2+4(1+\sigma\Gamma)(1-\alpha)+3(1-\alpha)\nonumber \\
&f^{(2)}=1,
\nonumber \\
&D^{(2)}(\Delta^{(2)},\sigma\Gamma,\alpha)=4\Delta^{(2)}(1+\sigma\Gamma)^3+2(1-\Delta^{(2)})\left[6(1+\sigma\Gamma)^2+2(1+\sigma\Gamma)\right.
\nonumber \\
&
\left.+2(1+\sigma\Gamma)(1-\alpha)+(1-\alpha)
 \right], 
 \label{F_dimer}
\end{align}
 where $\Delta^{(2)}=1-\displaystyle\frac{3}{2}\eta_I$.
For model~A  with the cations made of three  monomers,
\begin{align}	
&F_1^{(3)}(\sigma\Gamma,\alpha)=8(1+\sigma\Gamma)^2+6(1+\sigma\Gamma)-6(1+\sigma\Gamma)(1-\alpha)-3(1-\alpha) \nonumber \\
& F_2^{(3)}(\sigma\Gamma,\alpha)=16(1+\sigma\Gamma)^3+8(1+\sigma\Gamma)^2+6(1+\sigma\Gamma)+8(1+\sigma\Gamma)^2(1-\alpha) \nonumber \\
& +6(1+\sigma\Gamma)(1-\alpha)+3(1-\alpha),\nonumber \\
& f^{(3)}(\sigma\Gamma,\alpha)=3[1+2(1+\sigma\Gamma)],
	\nonumber \\
& D^{(3)}(\Delta^{(3)},\sigma\Gamma,\alpha)=32\Delta^{(3)}(1+\sigma\Gamma)^4+3(1-\Delta^{(3)})\left[32(1+\sigma\Gamma)^3+12(1+\sigma\Gamma)^2\right.
	\nonumber \\
&	\left.+2(1+\sigma\Gamma)+4(1+\sigma\Gamma)(1-\alpha)+8(1+\sigma\Gamma)^2(1-\alpha)\right.
	\left. +(1-\alpha) \right],\nonumber
  \label{F_trimer}
\end{align}
where $\Delta^{(3)}=1-2\eta_I$.

Again, if $\eta^B_{m_c}=0$ in (\ref{Gamma}), the equation for the RPM in the AMSA is recovered \cite{Bernard96,Jiang02,holovko2017effects}.
In order to obtain $\alpha$ and $\Gamma$, Eqs.~(\ref{MALEe}) and (\ref{Gamma}) should be solved self-consistently.
If we put $\alpha=1$ in (\ref{Gamma})-(\ref{eta_B}), that corresponds to the case of a complete ion dissociation ($K=0$ in Eq.~(\ref{MALEe})), the equation for $\Gamma$ transforms to the equation for the screening parameter in the MSA. 

The contribution to the free energy due to the formation of chain cations, $\beta f^{(ch)}$, is given by
\begin{equation}
	\beta f^{(ch)}=-\frac{\rho_{I}}{2}(m_c-1)\ln{\left[g^{(hs)}(\sigma^+)\right]},
	\label{f-ch}
\end{equation}
where $g^{(hs)}(\sigma^+)$ is the contact value of the pair distribution function of the HSs in the reference system.  

For the contribution from the electrostatic ion interaction, we use a simple interpolation scheme  (SIS) approximation 
introduced by Stell and Zhou~\cite{Zhou_SPM}. In the SIS approximation, we have
\begin{equation}
\beta f^{(el)}=-\frac{\rho_{I}}{T^*}\left(\frac{\sigma\Gamma^{(0)}}{1+\sigma\Gamma^{(0)}}+\frac{\eta_{B,m_c}^{(0)}\sigma^2}{\sum_{l=2}^{m_c}2^l(1+\sigma\Gamma^{(0)})^l}\right) +\frac{\left(\Gamma^{(0)} \right)^{3}}{3\pi},
\label{f_el}
\end{equation}
where $\Gamma^{(0)}=\Gamma\vert_{\alpha=1}$ and $\eta_{B,m_c}^{(0)}=\eta^B_{m_c}\vert_{\alpha=1}$. $\Gamma^{(0)}$ and $\eta_{B,m_c}^{(0)}$ can be found by putting $\alpha=1$ in (\ref{Gamma})-(\ref{eta_B}).

Now, several comments are in order. Eqs.~(\ref{K_gamma}) and (\ref{f-ch}) include the contact values of the pair distribution functions at zero charge $g_{12}^{hs}(\sigma^+)$ and $g^{(hs)}(\sigma^+)$, respectively. Within the framework of the AMSA theory, these contact values coincide with the corresponding contact values of the pair distribution functions of the reference system.  The reference system of  model~A is a HS fluid confined in a disordered HS matrix. In this case, using the results of \cite{Hvozd2024}, we get for $g^{(hs)}(\sigma^+)$
\begin{equation}
g^{(hs)}(\sigma^+)=\frac{1}{\phi_0-\eta}+\frac{3}{2}\frac{\left(k_{0}\eta_0+\eta\right)}{(\phi_0-\eta)^2}
	+\frac{\left(k_{0}\eta_0+\eta\right)^2}{2(\phi_0-\eta)^3},
\label{gij}
\end{equation}
where $k_{0}=\sigma/\sigma_0$ and $\eta$ is the volume fraction of all monomers that make up the HS fluid. 
We use  (\ref{gij}) for $g^{(hs)}(\sigma^+)$ entering Eq.~(\ref{f-ch}), i.e. 
 for the monomers of the cation. For the contact value of the pair distribution function between the charged monomers of the cation and the anion at zero charges, i.e., $g_{12}^{(hs)}(\sigma_{12}^+)$ in (\ref{K_gamma}) we supplement (\ref{gij}) by the correction  $\delta g_{12}^{(hs)}(\sigma_{12}^+)$ which arises in the so-called polymer Percus-Yevick ideal-chain approximation \cite{Kalyuzhnyi1998:2}. In the presence of the matrix, the correction has the form:
 \[
 \delta g_{12}^{(hs)}(\sigma_{12}^+)=-\frac{1}{4(\phi_0-\eta)}.
 \]
As a result,  we obtain for $g_{12}^{(hs)}(\sigma_{12}^+)$
\[
g_{12}^{(hs)}(\sigma_{12}^+)=g^{(hs)}(\sigma^+)+ \delta g_{12}^{(hs)}(\sigma_{12}^+).
\]

 Using (\ref{f-ass}), (\ref{f-ch}), and (\ref{f_el}), one can get  the corresponding contributions to the pressure $P$ and the chemical potential $\mu=(\mu_++\mu_-)$ from the standard thermodynamic relations
\[
P=-f+\rho_{I}\frac{\partial f}{\partial\rho_{I}}, \qquad
\frac{\rho_{I}}{2}\left(\mu_++\mu_-\right)=f+P.
\]
As a result, $\beta P^{\rm{(ass)}}$ and $\beta\mu^{\rm{(ass)}}$ are as follows:
 \begin{align}
&\beta P^{\rm{(ass)}}=-\frac{\rho_{I}}{2}(1-\alpha)\left(1+\rho_{I}\frac{\partial\ln K^{\gamma}}{\partial\rho_{I}}\right),
\label{P_ass} \\
&\beta\mu^{\rm{(ass)}}=\ln\alpha-\frac{\rho_{I}}{2}(1-\alpha)\frac{\partial\ln K^{\gamma}}{\partial\rho_{I}}.
\label{mu_ass}
 \end{align}
It is worth noting that the equations  (\ref{P_ass})-(\ref{mu_ass}) are formally similar to the expressions for $\beta P^{\rm{(mal)}}$ and $\beta\mu^{\rm{(mal)}}$ obtained for  the ions of spherical shape in \cite{holovko2017effects,holovko2017application}. However, the main difference of the above expressions is that the parameters $\alpha$, $K^{\gamma}$, and $\Gamma$  are determined by Eqs.~(\ref{MALEe}), (\ref{K_gamma}), and (\ref{Gamma}), respectively. These equations account for the non-spherical shape of the cations, as determined by the asymmetry parameter $\eta_{B,m_c}^{(0)}$.

 In a similar way, one gets analytical expressions for $\beta P^{\rm{(ch)}}$ and $\beta\mu^{\rm{(ch)}}$
 \begin{align}
 &\beta P^{\rm{(ch)}}=-\frac{\rho_{I}^2}{2}(m_c-1)\frac{\partial}{\partial\rho_{I}}\ln{\left[g^{(hs)}(\sigma^+)\right]},
 \label{P_ch} \\
&\beta\mu^{\rm{(ch)}}=-(m_c-1)\frac{\partial}{\partial\rho_{I}}\left[\rho_{I}\ln{\left[g^{(hs)}(\sigma^+)\right]}\right]
\label{mu_ch}
 \end{align}
 and for the electrostatic contributions $\beta P^{\rm{(el)}}$ and $\beta\mu^{\rm{(el)}}$ \cite{Bernard96,Kalyuzhnyi2018}
\begin{align}
	\label{P0_el}
&	\beta P^{(el)}=-\frac{(\Gamma^{(0)})^3}{3\pi}-\frac{2\beta e^2}{\pi\varepsilon}(\eta_{B,m_c}^{(0)})^2,\\
	\label{mu_el-CAL}
&	\beta\mu^{(el)}=-\frac{2}{T^*}\left(\frac{\sigma\Gamma^{(0)}}{1+\sigma\Gamma^{(0)}}+\frac{\eta_{B,m_c}^{(0)}\sigma^2}{\sum_{l=2}^{m_c}2^l(1+\sigma\Gamma^{(0)})^l}\right)-\frac{4\beta e^2}{\pi\varepsilon\rho_{I}}(\eta_{B,m_c}^{(0)})^2.
\end{align}

As a result, the contributions to the pressure and the chemical potential from the ionic subsystem can be presented as follows:
\begin{align}
&\beta\Delta P = \beta P^{(ass)}+\beta P^{(ch)}+\beta P^{(el)},
\label{Delata_p} \\
&\beta\Delta\mu = \beta\mu^{(ass)}+\beta\mu^{(ch)}+\beta\mu^{(el)},
\label{Delata_mu}
\end{align}
where the expressions for the corresponding addends are given by (\ref{P_ass})-(\ref{mu_el-CAL}). The contribution to the pressure and the chemical potential from the RS are given by Eqs.~(\ref{pressureCS1})-(\ref{muCS-1}).

For model~B, the contributions to the pressure and chemical potential from the ionic subsystem can be represented by Eqs.~(\ref{Delata_p})-(\ref{Delata_mu}) setting the terms $\beta P^{(ch)}$ and $\beta\mu^{(ch)}$ equal to zero. In addition, it should be kept in mind that the HSC lengths $L^*=1$ and $L^*=2$ correspond to the numbers of monomers in cation chain $m_c=2$ and $m_c=3$, respectively. It should be emphasised that the contributions to both the pressure $P$ and the chemical potential  $\mu=\mu_1+\mu_2$ from the RS, presented as a mixture of HSs and HSCs confined in a disordered HS matrix, are given by the expressions described in paragraph~\ref{par2-2-2} and Appendix~A.

\section{Results and discussion} 
The vapour-liquid phase transition of the models A and B (Table~\ref{table1}) in the bulk and in a disordered porous medium is studied 
using the theoretical approach presented in the previous section.
For this purpose, the expressions for chemical potential $\mu$  and pressure $P$  presented in subsections~\ref{subs1} and \ref{subs2} are used to find thermodynamic equilibrium between two phases 
coexisting at temperatures below the critical temperature of the vapour-liquid phase transition. 
The conditions of thermodynamic equilibrium are as follows:
\begin{align} \label{mu_P_eq}
\begin{split}
	&\mu(\rho_{I}^{(v)},\alpha^{(v)},\eta_{0},T)=\mu(\rho_{I}^{(l)},\alpha^{(l)},\eta_{0},T),
	\\ 
	&P(\rho_{I}^{(v)},\alpha^{(v)},\eta_{0},T)=P(\rho_{I}^{(l)},\alpha^{(l)},\eta_{0},T), 
\end{split} 
\end{align}    
where the superscripts ``$(v)$'' or ``$(l)$'' correspond to the vapour or liquid phases, respectively.
This set of equations is solved numerically for the ILs in a matrix of different packing fractions $\eta_0 = 0.0$, $0.05$, $0.1$ at given temperatures. The Newton-Raphson iterative algorithm is adapted to perform both outer and inner iterative procedures to determine the coexistence densities, $\rho_{I}^{(v)}$ and $\rho_{I}^{(l)}$, 
as well as the dissociation degrees, $\alpha^{(v)}$ and $\alpha^{(l)}$. The inner procedure solves the coupled equations \eqref{MALEe} and \eqref{Gamma} for $\Gamma$ and $\alpha$, starting with values corresponding to the limit of the full association. This is done for an IL in each phase, and the resulting $\Gamma$ and $\alpha$ are then used to calculate the chemical potential and pressure. These values are subsequently employed in the outer procedure to solve the set of equations~\eqref{mu_P_eq}.
The pairs of $\rho_{I}^{(v)}$ and $\rho_{I}^{(l)}$, as well as $\alpha^{(v)}$ and $\alpha^{(l)}$, obtained at corresponding temperatures $T$ are used to build phase diagrams for model~A and model~B.

\begin{figure}[!ht]
	\centering
	\includegraphics[width=0.4\linewidth]{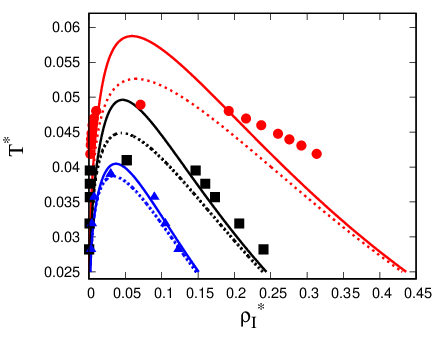}
	\includegraphics[width=0.4\linewidth]{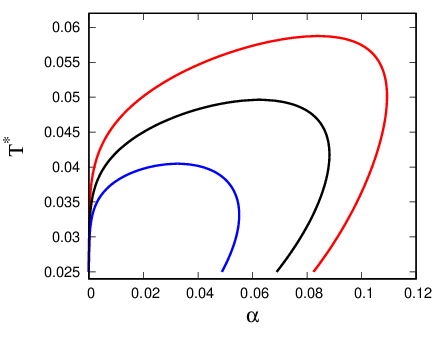}\\
	\caption{
		Vapour-liquid phase diagrams of model~A in the bulk presented in terms of $\rho_{I}^{*}-T^*$ (left panel) and $\alpha-T^*$ (right panel) coordinates for the cation chains of different lengths $m_c=1$ (RPM model) (red), $m_c=2$ (black) and $m_c=3$ (blue). Solid lines correspond to the results obtained by taking into account the partial association between cations and anions, while the dotted lines correspond to the approximation of fully associated ions.
		Symbols denote computer simulation results taken from~\cite{Kalyuzhnyi2018}. $T^*=k_B T\varepsilon\sigma/\rm{e}^2$, $\rho_{I}$ is the number density of ions, $\alpha$ is the fractions of free anions (cations),  $m_c$ is the number of monomers in the cation chain.}
	\label{Figure1} 
\end{figure}

First, we examine model~A, composed of cation chains and monomeric anions~(Table~\ref{table1}), where the cation chain length is set to $m_c=2$ or $m_c=3$. 
We begin by considering this model in the bulk ($\eta_0=0.0$). In the corresponding phase diagrams shown in Fig.~\ref{Figure1}, it is observed that increasing the cation chain length leads to a decrease in both the critical temperature and the critical density (see solid lines in the left panel of Fig.\ref{Figure1} and Appendix~B, Table~\ref{table2}), while simultaneously narrowing the overall phase coexistence region. 
This trend qualitatively aligns with the computer simulation results reported in \cite{Kalyuzhnyi2018} for the same model and with our recent theoretical findings obtained using the AMSA approximation for fully associated ions \cite{Hvozd2024} (see also dashed lines in Fig.~\ref{Figure1}). The decrease of the critical temperature with increasing cation elongation is also in qualitative correspondence with experimental data for a homologous series of imidazolium-based ionic liquids \cite{Rebelo2005}. The reduction in the critical temperature can be attributed to the screening effect caused by the neutral tails of the cations and to the excluded volume effect, which lowers the densities within the phase coexistence region. 
Incorporating the partial association calculated using the present theory, however, yields higher values of the critical temperature and the critical density compared to the case of the full association, and it simultaneously broadens the phase coexistence region. This effect arises from weakening the association between the cations and the anions, similar to observations for the RPM model~\cite{holovko2017effects}. 
Moreover, as the neutral chain of the cations increases in size, the association becomes more significant. This effect is illustrated in Fig.~\ref{Figure1} (right panel), where the dissociation degree $\alpha$, calculated along the coexistence curves, decreases with increasing $m_c$, indicating a strengthening of the association between cations and anions. 
It is worth mentioning that although the phase diagrams obtained for partially associated ions show somewhat poorer agreement with the simulation data compared to the approximation of fully associated ions, they result from a more accurate and systematic consideration of the MAL contribution to the thermodynamic properties of the ionic system. Furthermore, the association/dissociation phenomena can occur in ionic liquids near the phase coexistence region, especially close to the critical point~\cite{camp1999ion}, affecting the phase behaviour in such systems, as observed here for the bulk case. 
The same trends in the phase behaviour were also obtained by Cheng et al.~\cite{Cheng2024} for identical models of ionic liquid studied both in the bulk and a slit-like pore using classical density functional theory. Similar to our study, they employed the first-order thermodynamic perturbation theory of Wertheim to account for the chain connectivity of cations and the mean spherical approximation (MSA) to address electrostatic interactions, which was combined with an associative contribution resulting from short-range Coulomb interactions between oppositely charged monomers. 
However, the critical temperatures reported in their study (cf. Figure~4d in \cite{Cheng2024}) for the bulk case are significantly higher than our results and show poorer agreement with the computer simulations~\cite{Kalyuzhnyi2018}.

\begin{figure}[!ht]
	\centering
	\includegraphics[width=0.4\linewidth]{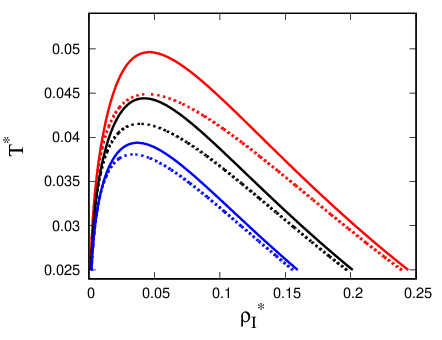}
	\includegraphics[width=0.4\linewidth]{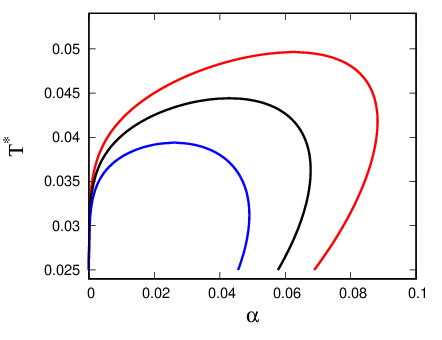}\\
	\includegraphics[width=0.4\linewidth]{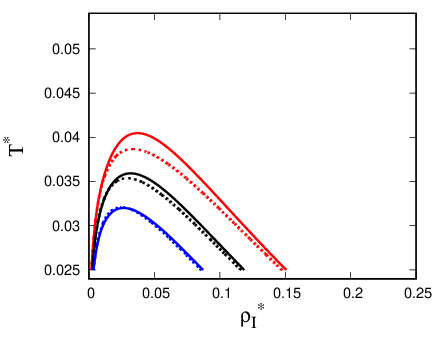}
	\includegraphics[width=0.4\linewidth]{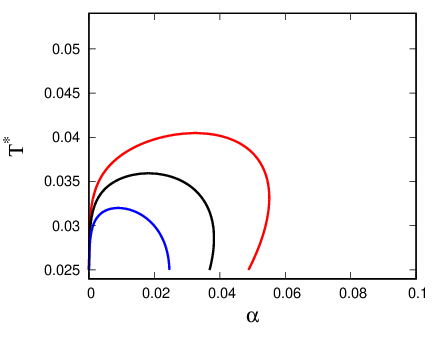}
	\caption{
		Vapour-liquid phase diagrams of model~A  in a disorder matrix presented in terms of $\rho_I^{*}-T^*$ (left panel) and $\alpha-T^*$ (right panel) coordinates for chain cations of lengths $m_c=2$ (upper panels) and $m_c=3$ (lower panels). The matrix packing fractions are $\eta_0=0.05$ (black lines) and $\eta_0=0.1$ (blue lines).
		The ionic liquid in the bulk ($\eta_0=0.0$) is shown for the reference (red lines).
		The solid and dotted lines correspond to partially and fully associated ions, respectively. The notations are the same as in Fig.~\ref{Figure1}.}
	\label{Figure2}
\end{figure}

Now, we consider model~A of an ionic liquid confined by a disordered matrix of hard sphere particles with packing fractions $\eta_0 = 0.05$ and $0.1$. Phase diagrams are calculated for model~A with cation lengths of $m_c = 2$ and $m_c = 3$ (Fig.\ref{Figure2}, left panel). As expected, the presence of fixed obstacles in the system lowers the critical temperature by weakening the overall interactions between ions, while the additional excluded volume effect shifts the phase coexistence region to lower density values and makes it narrower. An increase in the packing fraction of matrix particles, $\eta_0$, enhances this confinement effect, making the pores of the matrix smaller and the excluded volume greater. Thus, the association between oppositely charged ions becomes more significant within the matrix. Consequently, as the matrix packing fraction increases, the difference between the results for partially and fully associated ions diminishes. This difference becomes negligible when the cation length is $m_c=3$, and the matrix packing fraction is $\eta_0 = 0.1$ (blue lines in Fig.\ref{Figure2}, lower panel). 
Supporting evidence is provided in (Fig.\ref{Figure2}, right panel), where it is shown that the dissociation degree $\alpha$ obtained along the coexistence curves decreases significantly in the matrix, especially for longer cation chains, approaching the system of fully associated ions. 
It should be noted that the effect of confinement for model~A is similar to what was previously observed for an RPM fluid in a disordered matrix\cite{holovko2017effects, holovko2017application} and in a slit-like pore~\cite{Pizio004, Pizio04, Pizio05, Liu2020}. 
Additionally, it qualitatively agrees with the findings of \cite{Cheng2024} for the model with chain cations (equivalent to model~A) confined in a slit-like pore of different sizes (cf. Figures~4 and 5 in \cite{Cheng2024}). Similar to our study, they demonstrated that stronger confinement caused by a decrease in pore size leads to a decrease in both the critical temperature and critical density, along with a narrowing of the phase coexistence region. Conversely, an increase in pore size reduces the confinement effect, causing the phase behaviour of the ionic liquid to approach the bulk case. Comparing the results for the critical temperature in the bulk, it should be noted that our result is closer to the results from computer simulations than the value of $T_c$ reported in \cite{Cheng2024} (see Table~\ref{table2}).

\begin{figure*}[!ht]
	\centering
	\includegraphics[width=0.4\linewidth]{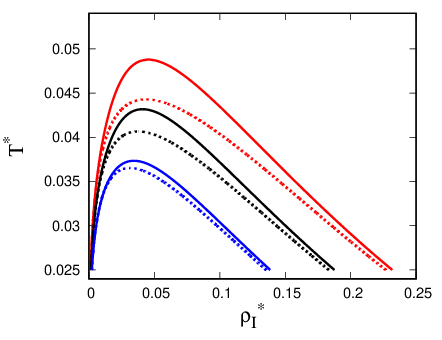}
	\includegraphics[width=0.4\linewidth]{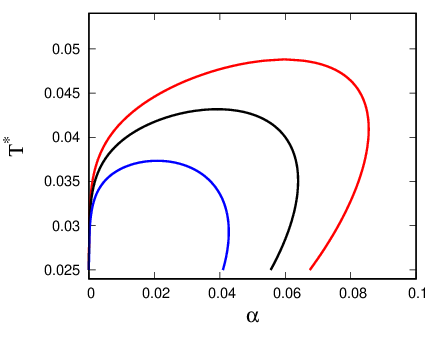}\\
	\includegraphics[width=0.4\linewidth]{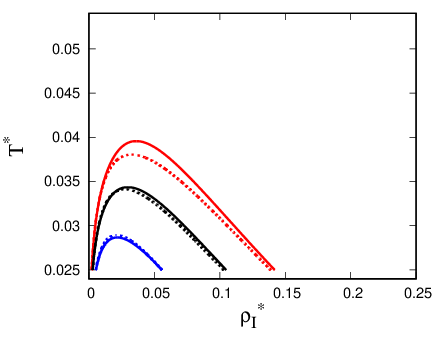}
	\includegraphics[width=0.4\linewidth]{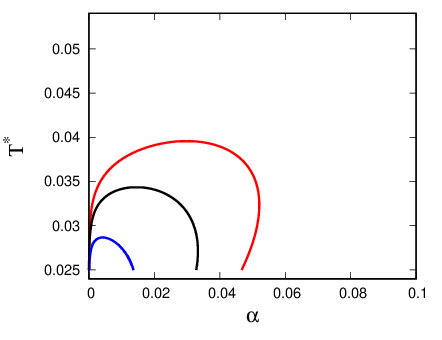}
    \caption{
		Vapour-liquid phase diagrams of model~B in a disorder matrix presented in terms of $\rho_I^{*}-T^*$ (left panel) and $\alpha-T^*$ (right panel) coordinates for spherocylinder cation of lengths $L^*=1$ (upper panels) and $L^*=2$ (lower panels). The matrix packing fractions are $\eta_0=0.05$ (black lines) and $\eta_0=0.1$ (blue lines).
        The ionic liquids in the bulk ($\eta_0=0.0$) are shown for the reference (red lines).
        The solid and dotted lines correspond to partially and fully associated ions, respectively. $L^*=L/\sigma$ and the other notations are the same as in Fig.~\ref{Figure1}.}
	\label{Figure3}
\end{figure*}

Next, we consider model~B, which consists of spherocylinder cations and monomeric anions~(Table~\ref{table1}), where the sizes of the cation molecules are taken to be the same as those for the cation chains in model~A. Specifically, the spherocylinder length $L^*=1$ corresponds to a chain length of $m_c=2$, and $L^*=2$ is equivalent to a chain length of $m_c=3$. 
Similar to the case of chain cations, Fig.~\ref{Figure3} shows the vapour-liquid phase diagrams for model~B in the bulk ($\eta_0=0.0$) and under matrix confinement with packing fractions of $0.05$ and $0.1$. It is clearly seen that the overall trend for model~B repeats the one observed for model~A. 
However, the confinement effect is more pronounced for spherocylinder cations, where it enhances the association between cations and anions, making the phase diagrams nearly indistinguishable in the partially and fully associated approximations even at a matrix packing fraction of $\eta_0=0.05$, 
particularly when the spherocylinder length is $L^*=2$. It may indicate the relatively minor presence of dissociated ions in the system under these conditions. This can also be inferred from the dissociation degree $\alpha$ obtained along the vapour-liquid coexistence curves (see Fig.~\ref{Figure3}, right panel). 
\begin{figure*}[!ht]
	\centering
	\includegraphics[width=0.4\linewidth]{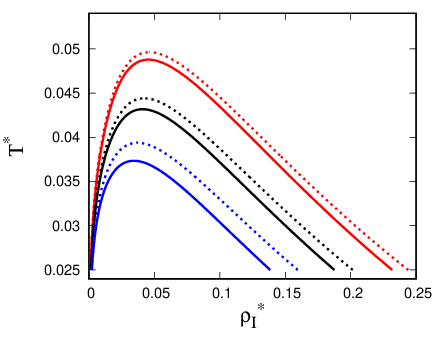}
	\includegraphics[width=0.4\linewidth]{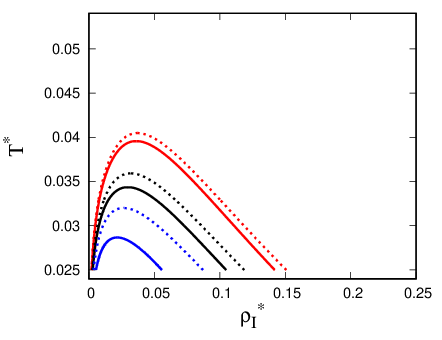}\\
	\caption{
		Comparison of vapour-liquid phase diagrams obtained for model~A and model~B in the bulk (red lines), and in a matrix with packing fraction $\eta_0=0.05$ (black lines) and $\eta_0=0.1$ (blue lines). 
		The dotted and solid lines correspond to the model~A and B, respectively.
		The systems with cations of sizes $m_c=2$ and $L^*=1$ are shown in the left panel, with cations of sizes $m_c=3$ and $L^*=2$ are shown in right panel. $L^*=L/\sigma$ and the other notations are the same as in Fig.~\ref{Figure1}.}
	\label{Figure4}
\end{figure*}
\noindent
Additionally, we compare our results for the phase diagrams of model~B with the coexistence curves obtained from Monte Carlo simulations 
in the bulk case~\cite{MartnBetancourt2009}. 
They considered the same model~B but with the lengths of spherocylinder cations ranging 
from $L^*=0.0$ to $1.0$ (cf. Figure~2 in ref.~\cite{MartnBetancourt2009}). 
It was shown that an increase in the length $L^*$ leads to a lowering of the critical temperature and a decrease in the critical density. This trend qualitatively coincides with the behaviour observed 
in our study~(Fig.\ref{Figure3}). However, for model~B ($L^*=1$) in the bulk (Fig.~\ref{Figure3}, upper panel), the critical temperature predicted by our theory is significantly higher compared to that reported by Mart{\'i}n-Betancourt et al. (cf. upside triangles in Figure~2 and Table~1 in ref.~\cite{MartnBetancourt2009} for $l^{*}=1$) even for the fully associated ions. On the other hand, the critical density estimated from our theory is in rather good agreement with their computer simulations. The critical parameters obtained for model~B are presented in Appendix~B, Table~\ref{table2}.

\begin{figure*}[!ht]
\centering
\includegraphics[width=0.4\linewidth]{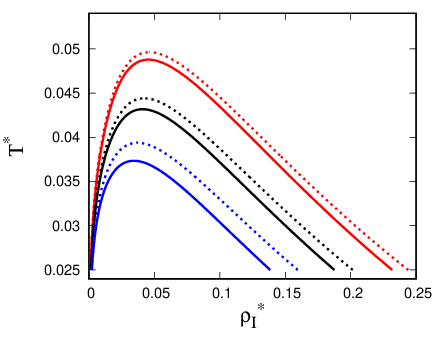}
\includegraphics[width=0.4\linewidth]{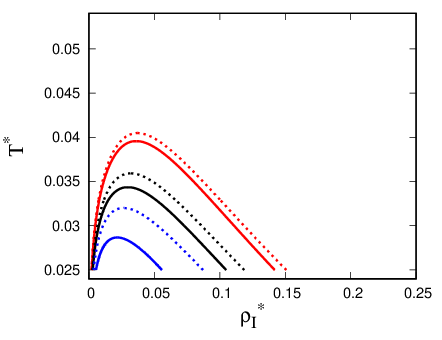}\\
\caption{
		Comparison of vapour-liquid phase diagrams obtained for model~A and model~B in the bulk (red lines), and in a matrix with packing fraction $\eta_0=0.05$ (black lines) and $\eta_0=0.1$ (blue lines). 
		The dotted and solid lines correspond to the model~A and B, respectively.
		The systems with cations of sizes $m_c=2$ and $L^*=1$ are shown in the left panel, with cations of sizes $m_c=3$ and $L^*=2$ are shown in right panel. $L^*=L/\sigma$ and the other notations are the same as in Fig.~\ref{Figure1}.}
\label{Figure4}
\end{figure*}

Finally, we compare the results obtained for models A and B for $\alpha\neq 0$  (partial association) to elucidate the effect of the shape of cation molecules 
on the phase behaviour of ionic liquids (the results for $\alpha=0$ is presented in Appendix~C, Fig.~\ref{Figure5}). In Fig.~\ref{Figure4} (left panel), we present the phase diagrams for the corresponding systems 
where the chain cations have a length of $m_c=2$, and the spherocylinder cations have a length of $L^*=1$. 
As it is seen, the phase coexistence region for model~B is shifted to lower temperatures and smaller densities compared to model~A. 
This shift is observed both in the bulk and in the matrix with packing fractions of $\eta_0=0.05$ and $\eta_0=0.1$. 
The same trend is observed for larger cations (Fig.~\ref{Figure4}, right panel), i.e., when the chain cations have a length of $m_c=3$ and the spherocylinder cations have a length of $L^*=2$. However, in this case, the phase coexistence region is located at significantly lower temperatures and densities 
compared to the system with smaller cations. Moreover, the difference between the results for models A and B becomes substantial when the matrix packing fraction is $\eta_0=0.1$. 

\section{Conclusions}
The effect of a disordered porous medium on the vapour-liquid phase behaviour of ILs was studied using the theoretical approach, which combines the SPT theory, Wertheim’s thermodynamic perturbation theory, and AMSA theory. The theoretical approach allows us to obtain analytical expressions for the thermodynamic functions of the IL models. We considered two models of molecular ILs, which differ in the shape of the molecular cation: one type of cation consisted of $m_c=2$ or $m_c=3$ tangentially bonded hard spheres of the equivalent diameters and with the charge placed on one of the terminal spheres; the other type of cation was modelled as a hard spherocylinder of the length $L^*=1$ or $L^*=2$ with the charge placed at the centre of one of its hemisphere caps. The anions were presented as negatively charged hard spheres. Therefore, the former model (model~A) described a system of ionic liquid containing cation molecules with flexible neutral tails, while the latter model (model~B) corresponded to an ionic liquid with cation molecules having rigid neutral tails. To compare the phase diagrams of these two models, we set the diameter of both spheres composing chain cations and spherocylinders to be equal. Furthermore, the lengths of the chains and spherocylinders were chosen to be the same when we examined the shape effect of the cations.

The vapour-liquid phase diagrams for these two models were obtained for the bulk phase and in a porous medium composed of a HS matrix with packing fractions of $\eta_0 = 0.05$ and $\eta_0 = 0.10$, corresponding to porosities of $\phi_0 = 0.95$ and $0.9$, respectively. The theory proposed in this study is an extension of the approach we developed earlier for an IL fluid with chain cations in the limit of fully associated ions. This time, we introduced the theory by accounting for the partial association of ions, thereby enabling the description of an arbitrary mixture of dimerised and free ions in the IL, which naturally arises in accordance with the mass action law. 
Additionally, the proposed theory is generalised to the case where cations are modelled as spherocylinders. To the best of our knowledge, the latter is the first theoretical attempt to describe such a model, particularly in a disordered porous medium.

Our main goal of this study was to examine the liquid-vapour phase behaviour of the model ILs affected by the following aspects: partial association of oppositely charged ions (association/dissociation phenomena), the shape of molecular cations (chain or spherocylider), and the presence of a disordered porous medium (confinement effect). Specifically, it was shown that taking into account the dissociation phenomena in the model ILs leads to an increase in both the critical temperature and critical density compared to the limit of fully dimerised ions, and it also widens the phase coexistence region. These trends are observed for both models. 
In a porous matrix, the dimerisation of oppositely charged ions becomes more significant than in the bulk, causing the phase diagrams to become closer and, at smaller porosities, even indistinguishable from the results obtained in the limit of fully associated ions. And this effect is more pronounced for the IL with the spherocylinder cations. 
Moreover, in both models, the critical temperature and critical density of ILs decrease simultaneously as the size of the cations increases, enhancing the effect of the porous medium, which affects the phase diagram of all considered ILs in a similar manner.

It was found that IL with spherocylinder cations have lower critical temperatures and smaller values of critical densities than those in IL systems with chain cations, and the phase coexistence region is narrower. This may result from the molecule rigidity of ILs in the case of larger cations, as well as their specific geometry, which differs significantly when comparing chains to spherocylinders. Our findings are in qualitative agreement with computer simulation for ILs with chain cations of size $m_c=2$ and $3$ reported in \cite{Kalyuzhnyi2018}, and for ILs with spherocylinder cations of sizes $L^*=0.0-1.0$ published in~\cite{MartnBetancourt2009}.  However, it is difficult to assert that this trend will persist with further increases in cation sizes (i.e. $m_c>3$ and $L^*>2$). Moreover, in the case of sufficiently long spherocylinder cations, orientational order may arise, and in addition to the vapor-liquid phase transition, an isotropic-nematic phase transition could occur. This complex phase behaviour will be the subject of a future investigation using the present theory combined with the Parsons-Lee approach~\cite{parsons1979nematic,lee1987numerical}, as in one of our previous studies~\cite{Hvozd2022}, where a mixture of RPM fluid with neutral spherocylinders was considered.

\section*{Acknowledgements} 
We gratefully acknowledge the financial support from the European Union external assistance instrument to fulfil Ukraine's obligations under the European Union Framework Program for Research and Innovation ``Horizon 2020'' (Agreement No~PH/16-2023). YK acknowledges financial support through the MSCA4Ukraine project (ID:~101101923), funded by the European Union.

\newpage
\section*{Appendix A. The reference system of model~B}
\label{AppendixA}
\renewcommand{\theequation}{A\arabic{equation}}
\setcounter{equation}{0}

Here, we present the parameters entering the expressions for the contributions $\beta P^{\rm{SPT2b3}^\star}$ and $\beta \mu_i^{\rm{SPT2b3}^\star}$ to the pressure and the chemical potentials of the reference system of model~B.

The probe particle porosity of species $1$ and $2$ are as follows \cite{holovko2017effects}:
\begin{align}
\label{probeporosity01}
\phi_{1}&=(1-\eta_0)\exp \bigg[-3k_{10}\left(1+k_{10}\right)\frac{\eta_0}{1-\eta_0}-\frac{9}{2} k_{10}^2 \frac{\eta_0^2}{(1-\eta_0)^2} \nonumber \\
&-k_{10}^3 \frac{\eta_0}{(1-\eta_0)^3}\left(1+\eta_0+\eta_0^2\right)\bigg],
\\
\label{probeporosity02}
\phi_{2}&=(1-\eta_0)\exp \bigg[-3 k_{20}\left(\frac{1}{2}(\gamma_2+1)+\gamma_2 k_{20}\right) \frac{\eta_0}{1-\eta_0}
-\frac{9}{2} k_{20}^2 \gamma_2 \frac{\eta_0^2}{(1-\eta_0)^2} \nonumber \\
&-k_{20}^3\frac{3\gamma_2-1}{2} \frac{\eta_0}{(1-\eta_0)^3} \left(1+\eta_0+\eta_0^2\right)\bigg],
\end{align}
\begin{align*}
k_{10}=k_{20}=\sigma/\sigma_0, \quad
\gamma_2=1+\frac{L}{\sigma}.
\end{align*}
Using (\ref{smallphip}) and (\ref{probeporosity01})-(\ref{probeporosity02}), one arrives at the new expression for $\phi^\star=\frac{\phi_0\phi}{\phi_0-\phi}\ln\frac{\phi_0}{\phi}$.

$A$ and $B$  are given by:
\begin{equation}
\label{Ap}
A=\frac{1}{2}(a_1+a_2), \qquad
B=\frac{1}{2}(b_1+b_2),
\end{equation}
where the coefficients $a_1$, $b_1$, $a_2$, $b_2$ have the form:
\begin{align}
a_1&=6\frac{\eta_1}{\eta}+\left[\frac{1}{k_1}\frac{6\gamma_2}{3\gamma_2-1}
+\frac{1}{k_1^2}\frac{3(\gamma_2+1)}{3\gamma_2-1}\right]\frac{\eta_2}{\eta}
-\frac{p'_0}{\phi_0}\left(3\frac{\eta_1}{\eta}+\frac{1}{k_1}\frac{6\gamma_2}{3\gamma_2-1}\frac{\eta_2}{\eta}\right)
\nonumber \\
&-\frac{p'_0}{\phi_0}+\left(\frac{p'_0}{\phi_0}\right)^2-\frac{1}{2}\frac{p''_0}{\phi_0}, \label{a1p}
\\
b_1&=\frac{1}{2}\bigg(3\frac{\eta_1}{\eta}+\frac{1}{k_1}\frac{6\gamma_2}{3\gamma_2-1}\frac{\eta_2}{\eta}
-\frac{p'_0}{\phi_0}\bigg)^2
\label{b1p}
\end{align}
and
\begin{align}
a_2&=\left[3k_1(1+k_1)+\frac{3}{4}s_1(1+2k_1)\right]\frac{\eta_1}{\eta}
+\left[6+\frac{6(\gamma_2-1)^2\tau(f)}{3\gamma_2-1}\right]\frac{\eta_2}{\eta} \nonumber \\
&-\frac{p'_{0\alpha}}{\phi_0}\left(1+3k_1\frac{\eta_1}{\eta}+\frac{6\gamma_2}{3\gamma_2-1}\frac{\eta_2}{\eta}\right)
\nonumber \\
&-\frac{p'_{0\lambda}}{\phi_0}\left[1+\left(3k_1+\frac{3}{4}s_1\right)\frac{\eta_1}{\eta}
+\left(3+\frac{3(\gamma_2-1)^2\tau(f)}{3\gamma_2-1}\right)\frac{\eta_2}{\eta}\right] \nonumber \\
&+2\frac{p'_{0\alpha} p'_{0\lambda}}{\phi_0^2}+\left(\frac{p'_{0\lambda}}{\phi_0}\right)^2
-\frac{p''_{0\alpha\lambda}}{\phi_0}-\frac{1}{2}\frac{p''_{0\lambda\lambda}}{\phi_0}, \label{a2p}
\\
b_2&=\left[\left(\frac{3}{4}s_1+\frac{3}{2}k_1\right)\frac{\eta_1}{\eta}
+\left(\frac{3(2\gamma_2-1)}{3\gamma_2-1}
+\frac{3(\gamma_2-1)^2\tau(f)}{3\gamma_2-1}\right)\frac{\eta_2}{\eta}
\right.
\nonumber \\
&
\left.
-\frac{p'_{0\alpha}}{\phi_0}-\frac{1}{2}\frac{p'_{0\lambda}}{\phi_0}\right]  \left(3k_1\frac{\eta_1}{\eta}+\frac{6\gamma_2}{3\gamma_2-1}\frac{\eta_2}{\eta}-\frac{p'_{0\lambda}}{\phi_0}\right).
\label{b2p}
\end{align}
The  notations in Eqs.~(\ref{b1p})-(\ref{b2p}) are as follows: 
$s_1=2L/\sigma$, 
$p'_0=-3\eta_{0}k_{10}$, $p''_0=-6\eta_{0}k_{10}^2$, $p'_{0\alpha}=-\frac{3}{4}\eta_{0}s_{0}$, $p'_{0\lambda}=-3\eta_{0}k_{20}$, $p''_{0\alpha\lambda}=-\frac{3}{2}\eta_{0}s_{0}k_{20}$, $p''_{0\lambda\lambda}=-6\eta_{0}s_{0}k_{20}^2$, where
$k_{10}=k_{20}=\sigma/\sigma_0$,  
$s_0=2L/\sigma_0$. For our model,  $k_1=\sigma_2/\sigma_1=1$. In addition, we do not take into account the orientation contribution for the HSCs with sufficiently short lengths $L$ and therefore put $\tau(f)=1$ in Eqs.~(\ref{a2p})-(\ref{b2p}).

Putting $L=0$, $\gamma=1$, $s_0=0$, $s_1=0$, $V_{1}=V_{2}=\pi\sigma^3/6$, and $\phi_1=\phi_2=\phi$ in the above formulas we obtain the corresponding expressions for the RS of model~A.

\newpage
\section*{Appendix B. The critical parameters of model~A and model~B in the bulk and confined in a disordered matrix}
\label{AppendixB}
\setcounter{table}{0}
\renewcommand{\tablename}{Table}
\renewcommand{\thetable}{B\arabic{table}}
\begin{table*}[h]
	\begin{center}
	\caption{Critical parameters of the ionic liquid models studied in this work. The notations $A_1$ and $A_2$  correspond to two versions of model~A with $m_c=2$ and $m_c=3$, respectively, while the notations $B_1$ and $B_2$ correspond to two versions of model~B with $L^*=1$ and $L^*=2$ ($L^*=L/\sigma$). For each $A_i$ and $B_i$, the results obtained in the limit of the full association are labelled as ``full'', and the results for the partial association are labelled as ``partial''. }
\begin{tabular}{|l|c|c|c|c|c|c|c|c|c|}
	\hline 
	\textbf{Models} & \multicolumn{3}{c|}{$\eta_0=0$}  &  \multicolumn{3}{c|}{$\eta_0=0.05$} &  \multicolumn{3}{c|}{$\eta_0=0.1$} \\
    \cline{2-10}
	 \textbf{of ionic liquids} &  $\rho^*_{cr}$ & $T^*_{cr}$ & $\alpha_{cr}$ & $\rho^*_{cr}$ & $T^*_{cr}$ & $\alpha_{cr}$ & $\rho^*_{cr}$ & $T^*_{cr}$ & $\alpha_{cr}$ \\
	\hline\hline
	\textbf{A ($m_c=2$)} -- full & $0.0447$ & $0.0449$ & 0 & $0.0390$ & $0.0415$ & 0 & $0.0336$ & $0.0380$ & 0 \\
	\hline
	\textbf{A ($m_c=2$)} - partial & $0.0462$ & $0.0496$ & $0.0624$ & $0.0422$ & $0.0444$ & $0.0429$ & $0.0370$ & $0.0394$ & $0.0261$ \\
	\hline
	\textbf{A ($m_c=3$)} -- full & $0.0335$    & $0.0387$ & 0    & $0.0288$ & $0.0354$ & 0 & $0,0249$ & $0.0320$ & 0 \\
	\hline
	\textbf{A ($m_c=3$)} -- partial & $0.0371$ & $0.0405$ & $0.0325$ & $0.0319$ & $0.0359$ & $0.0182$ & $0,0265$ & $0.0320$ & $0.0089$ \\
	\hline
	\textbf{B ($L^*=1$)} -- full & $0.0436$ & $0.0443$ & 0 & $0.0374$ & $0.0407$ & 0 & $0,0312$ & $0.0365$ & 0 \\
	\hline
	\textbf{B ($L^*=1$)} -- partial & $0.0455$ & $0.0488$ & $0.0597$ & $0.0410$ & $0.0432$ & $0.0392$ & $0.0344$ & $0.0373$ & $0.0208$ \\
	\hline
	\textbf{B ($L^*=2$)} -- full & $0.0326$    & $0.0380$ & 0  & $0.0274$ & $0.0341$ & 0 & $0.0210$ & $0.0289$ & 0 \\
	\hline
	\textbf{B ($L^*=2$)} -- partial & $0.0361$ & $0.0396$ & $0.0298$ & $0.0298$ & $0.0343$ & $0.0145$ & $0.0216$ & $0.0287$ & $0.0042$ \\
	\hline
\end{tabular}
\label{table2}
\end{center}
\end{table*}


\newpage
\section*{Appendix C. Vapour--liquid phase diagrams of model~A and model~B in the limit of full ionic association }
\label{AppendixC}
\setcounter{figure}{0}
\renewcommand{\figurename}{Fig.}
\renewcommand{\thefigure}{C\arabic{figure}}

\begin{figure*}[h]
\centering
	\includegraphics[width=0.4\linewidth]{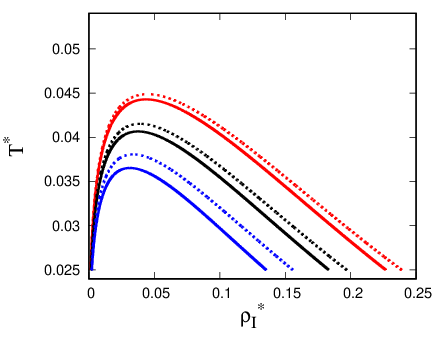}
	\includegraphics[width=0.4\linewidth]{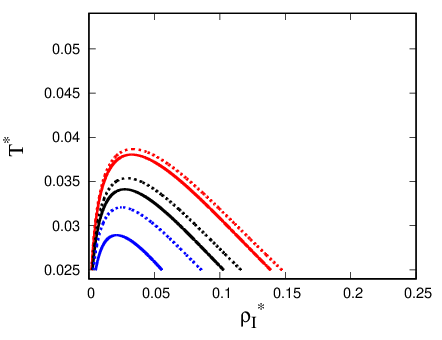}\\
	\caption{Comparison of vapour-liquid phase diagrams obtained for  model~A and model~B in the bulk (red lines), and in a matrix with packing fraction $\eta_0=0.05$ (black lines) and $\eta_0=0.1$ (blue lines) in the limit of full ionic association. 
		The dotted and solid lines correspond to the model~A and B, respectively.
		The systems with cations of sizes $m_c=2$ and $L^*=1$ are shown in the left panel, with cations of sizes $m_c=3$ and $L^*=2$ are shown in the right panel. $L^*=L/\sigma$ and the other notations are the same as in Fig.~\ref{Figure1}.}
	\label{Figure5}
\end{figure*}

\section*{Data Availability}
Data will be made available on request.

\newpage

\renewcommand{\refname}{References}
\def\bibsection{\section*{\refname}}
\bibliographystyle{elsarticle-num}
\bibliography{ionhscass}

\begin{thebibliography}{10}
\expandafter\ifx\csname url\endcsname\relax
  \def\url#1{\texttt{#1}}\fi
\expandafter\ifx\csname urlprefix\endcsname\relax\def\urlprefix{URL }\fi
\expandafter\ifx\csname href\endcsname\relax
  \def\href#1#2{#2} \def\path#1{#1}\fi

\bibitem{Seddon1997}
K.~R. Seddon, Ionic liquids for clean technology, Journal of Chemical
  Technology \& Biotechnology 68~(4) (1997) 351–356.
\newblock \href
  {http://dx.doi.org/10.1002/(sici)1097-4660(199704)68:4<351::aid-jctb613>3.0.co;2-4}
  {\path{doi:10.1002/(sici)1097-4660(199704)68:4<351::aid-jctb613>3.0.co;2-4}}.

\bibitem{Fedorov2014}
M.~V. Fedorov, A.~A. Kornyshev, Ionic liquids at electrified interfaces,
  Chemical Reviews 114~(5) (2014) 2978–3036.
\newblock \href {http://dx.doi.org/10.1021/cr400374x}
  {\path{doi:10.1021/cr400374x}}.

\bibitem{Eftekhari2017}
A.~Eftekhari, Supercapacitors utilising ionic liquids, Energy Storage Materials
  9 (2017) 47–69.
\newblock \href {http://dx.doi.org/10.1016/j.ensm.2017.06.009}
  {\path{doi:10.1016/j.ensm.2017.06.009}}.

\bibitem{Jeanmairet2022}
G.~Jeanmairet, B.~Rotenberg, M.~Salanne, Microscopic simulations of
  electrochemical double-layer capacitors, Chemical Reviews 122~(12) (2022)
  10860–10898.
\newblock \href {http://dx.doi.org/10.1021/acs.chemrev.1c00925}
  {\path{doi:10.1021/acs.chemrev.1c00925}}.

\bibitem{Wu2022}
J.~Wu, Understanding the electric double-layer structure, capacitance, and
  charging dynamics, {\it Chem. Rev.} 122~(12) (2022) 10821--10859.
\newblock \href {http://dx.doi.org/10.1021/acs.chemrev.2c00097}
  {\path{doi:10.1021/acs.chemrev.2c00097}}.

\bibitem{Kondrat2023}
S.~Kondrat, G.~Feng, F.~Bresme, M.~Urbakh, A.~A. Kornyshev, Theory and
  simulations of ionic liquids in nanoconfinement, Chem. Rev. 123~(10) (2023)
  6668--6715.
\newblock \href {http://dx.doi.org/10.1021/acs.chemrev.2c00728}
  {\path{doi:10.1021/acs.chemrev.2c00728}}.

\bibitem{LeBideau2011}
J.~Le~Bideau, L.~Viau, A.~Vioux, Ionogels, ionic liquid based hybrid materials,
  Chem. Soc. Rev. 40~(2) (2011) 907–925.
\newblock \href {http://dx.doi.org/10.1039/c0cs00059k}
  {\path{doi:10.1039/c0cs00059k}}.

\bibitem{singh2014ionic}
M.~P. Singh, R.~K. Singh, S.~Chandra, Ionic liquids confined in porous
  matrices: physicochemical properties and applications, Prog. Mater. Sci. 64
  (2014) 73--120.
\newblock \href {http://dx.doi.org/10.1016/j.pmatsci.2014.03.001}
  {\path{doi:10.1016/j.pmatsci.2014.03.001}}.

\bibitem{Zhang2016}
S.~Zhang, J.~Zhang, Y.~Zhang, Y.~Deng, Nanoconfined ionic liquids, Chem. Rev.
  117~(10) (2017) 6755–6833.
\newblock \href {http://dx.doi.org/10.1021/acs.chemrev.6b00509}
  {\path{doi:10.1021/acs.chemrev.6b00509}}.

\bibitem{Cheng2024}
J.~Cheng, J.~Xu, S.~Wang, X.~Chen, C.~Lian, H.~Liu, Phase behavior of ionic
  fluid in charged confinement: An associating polymer density functional
  theory study, AIChE Journal 70~(9) (2024) e18496.
\newblock \href {http://dx.doi.org/10.1002/aic.18496}
  {\path{doi:10.1002/aic.18496}}.

\bibitem{Pizio004}
O.~Pizio, A.~Patrykiejew, S.~Soko{\l}owski, Phase behavior of ionic fluids in
  slitlike pores: A density functional approach for the restricted primitive
  model, J. Chem. Phys. 121~(23) (2004) 11957--11964.
\newblock \href {http://dx.doi.org/10.1063/1.1818677}
  {\path{doi:10.1063/1.1818677}}.

\bibitem{Pizio05}
O.~Pizio, S.~Soko{\l}owski, Phase behavior of the restricted primitive model of
  ionic fluids with association in slitlike pores. density-functional approach,
  J. Chem. Phys. 122~(14) (2005) 144707.
\newblock \href {http://dx.doi.org/10.1063/1.1883165}
  {\path{doi:10.1063/1.1883165}}.

\bibitem{Liu2018}
K.~Liu, P.~Zhang, J.~Wu, Does capillary evaporation limit the accessibility of
  nonaqueous electrolytes to the ultrasmall pores of carbon electrodes?, J.
  Chem. Phys. 149~(23) (2018) 234708.
\newblock \href {http://dx.doi.org/10.1063/1.5064360}
  {\path{doi:10.1063/1.5064360}}.

\bibitem{Liu2020}
K.~Liu, J.~Wu, Wettability of ultra-small pores of carbon electrodes by
  size-asymmetric ionic fluids, The Journal of Chemical Physics 152~(5).
\newblock \href {http://dx.doi.org/10.1063/1.5131450}
  {\path{doi:10.1063/1.5131450}}.

\bibitem{Loubet2016}
M.~M. B.~Loubet, J.~Palmeri, A variational approach to the liquid-vapor phase
  transition for hardcore ions in the bulk and in nanopores, J. Chem. Phys.
  145~(4) (2016) 044107.
\newblock \href {http://dx.doi.org/10.1063/1.4959034}
  {\path{doi:10.1063/1.4959034}}.

\bibitem{Malvaldi2007}
M.~Malvaldi, C.~Chiappe, From molten salts to ionic liquids: effect of ion
  asymmetry and charge distribution, J. Phys.: Condens. Matter 20~(3) (2007)
  035108.
\newblock \href {http://dx.doi.org/10.1088/0953-8984/20/03/035108}
  {\path{doi:10.1088/0953-8984/20/03/035108}}.

\bibitem{Spohr2009}
H.~V. Spohr, G.~N. Patey, Structural and dynamical properties of ionic liquids:
  The influence of charge location, J. Chem. Phys. 130~(10) (2009) 104506.
\newblock \href {http://dx.doi.org/10.1063/1.3078381}
  {\path{doi:10.1063/1.3078381}}.

\bibitem{MartnBetancourt2009}
M.~Mart{\'{\i}}n-Betancourt, J.~M. Romero-Enrique, L.~F. Rull, Liquid-vapor
  coexistence in a primitive model for a room-temperature ionic liquid, J.
  Phys. Chem. B 113~(27) (2009) 9046--9049.
\newblock \href {http://dx.doi.org/10.1021/jp903709k}
  {\path{doi:10.1021/jp903709k}}.

\bibitem{Fedorov2010}
M.~Fedorov, N.~Georgi, A.~Kornyshev, Double layer in ionic liquids: The nature
  of the camel shape of capacitance, Electrochem. Commun. 12~(2) (2010)
  296--299.
\newblock \href {http://dx.doi.org/10.1016/j.elecom.2009.12.019}
  {\path{doi:10.1016/j.elecom.2009.12.019}}.

\bibitem{Wu2011}
J.~Wu, T.~Jiang, D.~Jiang, Z.~Jin, D.~Henderson, A classical density functional
  theory for interfacial layering of ionic liquids, Soft Matter 7~(23) (2011)
  11222.
\newblock \href {http://dx.doi.org/10.1039/c1sm06089a}
  {\path{doi:10.1039/c1sm06089a}}.

\bibitem{Ganzenmller2011}
G.~Ganzenm\"{u}ller, P.~Camp, Phase behaviour and dynamics in primitive models
  of molecular ionic liquids, Condens. Matter Phys. 14~(3) (2011) 33602.
\newblock \href {http://dx.doi.org/10.5488/cmp.14.33602}
  {\path{doi:10.5488/cmp.14.33602}}.

\bibitem{Lindenberg2014}
E.~K. Lindenberg, G.~N. Patey, How distributed charge reduces the melting
  points of model ionic salts, J. Chem. Phys. 140~(10) (2014) 104504.
\newblock \href {http://dx.doi.org/10.1063/1.4867275}
  {\path{doi:10.1063/1.4867275}}.

\bibitem{Lindenberg2015}
E.~K. Lindenberg, G.~N. Patey, Melting point trends and solid phase behaviors
  of model salts with ion size asymmetry and distributed cation charge, J.
  Chem. Phys. 143~(2) (2015) 024508.
\newblock \href {http://dx.doi.org/10.1063/1.4923344}
  {\path{doi:10.1063/1.4923344}}.

\bibitem{Guzmn2015}
O.~Guzm\'{a}n, J.~E. Ramos~Lara, F.~del R\'{i}o, Liquid–vapor equilibria of
  ionic liquids from a {SAFT} equation of state with explicit electrostatic
  free energy contributions, J. Phys. Chem. B 119~(18) (2015) 5864--5872.
\newblock \href {http://dx.doi.org/10.1021/jp511571h}
  {\path{doi:10.1021/jp511571h}}.

\bibitem{SilvestreAlcantara2016}
W.~Silvestre-Alcantara, L.~Bhuiyan, S.~Lamperski, M.~Kaja, D.~Henderson, Double
  layer for hard spheres with an off-center charge, Condens. Matter Phys.
  19~(1) (2016) 13603.
\newblock \href {http://dx.doi.org/10.5488/cmp.19.13603}
  {\path{doi:10.5488/cmp.19.13603}}.

\bibitem{Lu2016}
H.~Lu, B.~Li, S.~Nordholm, C.~E. Woodward, J.~Forsman, Ion pairing and phase
  behaviour of an asymmetric restricted primitive model of ionic liquids, J.
  Chem. Phys. 145~(23) (2016) 234510.
\newblock \href {http://dx.doi.org/10.1063/1.4972214}
  {\path{doi:10.1063/1.4972214}}.

\bibitem{Kalyuzhnyi2018}
Y.~Kalyuzhnyi, J.~Re{\v{s}}{\v{c}}i{\v{c}}, M.~Holovko, P.~Cummings, Primitive
  models of room temperature ionic liquids. liquid-gas phase~coexistence, J.
  Mol. Liq. 270 (2018) 7--13.
\newblock \href {http://dx.doi.org/10.1016/j.molliq.2018.01.109}
  {\path{doi:10.1016/j.molliq.2018.01.109}}.

\bibitem{Yao2023}
B.~Yao, M.~Paluch, J.~Paturej, S.~McLaughlin, A.~McGrogan, M.~Swadzba-Kwasny,
  J.~Shen, B.~Ruta, M.~Rosenthal, J.~Liu, D.~Kruk, Z.~Wojnarowska,
  Self-assembled nanostructures in aprotic ionic liquids facilitate charge
  transport at elevated pressure, ACS Appl. Mater. Interfaces 15~(33) (2023)
  39417–39425.
\newblock \href {http://dx.doi.org/10.1021/acsami.3c08606}
  {\path{doi:10.1021/acsami.3c08606}}.

\bibitem{holovko2009highly}
M.~Holovko, W.~Dong, A highly accurate and analytic equation of state for a
  hard sphere fluid in random porous media, J. Phys. Chem. B 113~(18) (2009)
  6360--6365.
\newblock \href {http://dx.doi.org/10.1021/jp809706n}
  {\path{doi:10.1021/jp809706n}}.

\bibitem{chen2009comment}
W.~Chen, W.~Dong, M.~Holovko, X.~S. Chen, Comment on ``{A} highly accurate and
  analytic equation of state for a hard sphere fluid in random porous media'',
  J.~Phys.~Chem.~B 114~(2) (2010) 1225.
\newblock \href {http://dx.doi.org/10.1021/jp9106603}
  {\path{doi:10.1021/jp9106603}}.

\bibitem{patsahan2011fluids}
T.~Patsahan, M.~Holovko, W.~Dong, Fluids in porous media. {III}. scaled
  particle theory, J. Chem. Phys. 134~(7) (2011) 074503: 1--11.
\newblock \href {http://dx.doi.org/10.1063/1.3532546}
  {\path{doi:10.1063/1.3532546}}.

\bibitem{holovko2012fluids}
M.~Holovko, T.~Patsahan, W.~Dong, Fluids in random porous media: Scaled
  particle theory, Pure Appl. Chem. 85~(1) (2013) 115--133.
\newblock \href {http://dx.doi.org/10.1351/pac-con-12-05-06}
  {\path{doi:10.1351/pac-con-12-05-06}}.

\bibitem{holovko2012one}
M.~F. Holovko, T.~Patsahan, W.~Dong, One-dimensional hard rod fluid in a
  disordered porous medium: Scaled particle theory, Condens. Matter Phys.
  15~(2) (2012) 23607: 1--13.
\newblock \href {http://dx.doi.org/10.5488/cmp.15.23607}
  {\path{doi:10.5488/cmp.15.23607}}.

\bibitem{holovko2017improvement}
M.~Holovko, T.~Patsahan, W.~Dong, On the improvement of {SPT2} approach in the
  theory of a hard sphere fluid in disordered porous media, Condens. Matter
  Phys. 20~(3) (2017) 33602: 1--14.
\newblock \href {http://dx.doi.org/10.5488/cmp.20.33602}
  {\path{doi:10.5488/cmp.20.33602}}.

\bibitem{chen2016scaled}
W.~Chen, S.~L. Zhao, M.~Holovko, X.~S. Chen, W.~Dong, Scaled particle theory
  for multicomponent hard sphere fluids confined in random porous media,
  J.~Phys.~Chem.~B 120~(24) (2016) 5491--5504.
\newblock \href {http://dx.doi.org/10.1021/acs.jpcb.6b02957}
  {\path{doi:10.1021/acs.jpcb.6b02957}}.

\bibitem{holovko2017isotropic}
M.~F. Holovko, M.~V. Hvozd, Isotropic-nematic transition in a mixture of hard
  spheres and hard spherocylinders: Scaled particle theory description,
  Condens. Matter Phys. 20 (2017) 43501.
\newblock \href {http://dx.doi.org/10.5488/cmp.20.43501}
  {\path{doi:10.5488/cmp.20.43501}}.

\bibitem{Hvozd2018}
M.~Hvozd, T.~Patsahan, M.~Holovko, Isotropic{\textendash}nematic transition and
  demixing behavior in binary mixtures of hard spheres and hard spherocylinders
  confined in a disordered porous medium: Scaled particle theory, J. Phys.
  Chem. B 122~(21) (2018) 5534--5546.
\newblock \href {http://dx.doi.org/10.1021/acs.jpcb.7b11834}
  {\path{doi:10.1021/acs.jpcb.7b11834}}.

\bibitem{holovko2016vapour}
M.~F. Holovko, O.~Patsahan, T.~Patsahan, Vapour-liquid phase diagram for an
  ionic fluid in a random porous medium, J. Phys.: Condens. Matter 28~(24)
  (2016) 414003.
\newblock \href {http://dx.doi.org/10.1088/0953-8984/28/41/414003}
  {\path{doi:10.1088/0953-8984/28/41/414003}}.

\bibitem{HolPatPat17}
M.~Holovko, T.~Patsahan, O.~Patsahan, Effects of disordered porous media on the
  vapour-liquid phase equilibrium in ionic fluids: application of the
  association concept, J. Mol. Liq. 228 (2017) 215--223.
\newblock \href {http://dx.doi.org/10.1016/j.molliq.2016.10.045}
  {\path{doi:10.1016/j.molliq.2016.10.045}}.

\bibitem{holovko2017application}
M.~F. Holovko, T.~M. Patsahan, O.~V. Patsahan, Application of the ionic
  association concept to the study of the phase behaviour of size-asymmetric
  ionic fluids in disordered porous media, J. Mol. Liq. 235 (2017) 53--59.
\newblock \href {http://dx.doi.org/10.1016/j.molliq.2016.11.030}
  {\path{doi:10.1016/j.molliq.2016.11.030}}.

\bibitem{patsahan2018vapor}
O.~V. Patsahan, T.~M. Patsahan, M.~F. Holovko, Vapor-liquid phase behavior of a
  size-asymmetric model of ionic fluids confined in a disordered matrix: {T}he
  collective-variables-based approach, Phys. Rev. E 97~(2) (2018) 022109.
\newblock \href {http://dx.doi.org/10.1103/physreve.97.022109}
  {\path{doi:10.1103/physreve.97.022109}}.

\bibitem{Hvozd2019}
M.~Hvozd, T.~Patsahan, O.~Patsahan, M.~Holovko, Fluid-fluid phase behaviour in
  the explicit solvent ionic model: Hard spherocylinder solvent molecules, J.
  Mol. Liq. 285 (2019) 244--251.
\newblock \href {http://dx.doi.org/10.1016/j.molliq.2019.03.171}
  {\path{doi:10.1016/j.molliq.2019.03.171}}.

\bibitem{Hvozd2022}
M.~Hvozd, O.~Patsahan, T.~Patsahan, M.~Holovko, Fluid-fluid phase behaviour in
  the explicit hard spherocylinder solvent ionic model confined in a disordered
  porous medium, J. Mol. Liq. 346 (2022) 117888.
\newblock \href {http://dx.doi.org/10.1016/j.molliq.2021.117888}
  {\path{doi:10.1016/j.molliq.2021.117888}}.

\bibitem{Hvozd2024}
T.~Hvozd, T.~Patsahan, Y.~Kalyuzhnyi, O.~Patsahan, M.~Holovko, Vapour-liquid
  phase behaviour of primitive models of ionic liquids confined in disordered
  porous media, Condens. Matter Phys. 27~(2) (2024) 23602.
\newblock \href {http://dx.doi.org/10.5488/cmp.27.23602}
  {\path{doi:10.5488/cmp.27.23602}}.

\bibitem{holovko2017effects}
M.~Holovko, T.~Patsahan, O.~Patsahan, Effects of disordered porous media on the
  vapour-liquid phase equilibrium in ionic fluids: application of the
  association concept, J. Mol. Liq. 228 (2017) 215--223.
\newblock \href {http://dx.doi.org/10.1016/j.molliq.2016.10.045}
  {\path{doi:10.1016/j.molliq.2016.10.045}}.

\bibitem{Carnahan_1969}
N.~F. Carnahan, K.~E. Starling, Equation of state for nonattracting rigid
  spheres, J. Chem. Phys. 51~(2) (1969) 635--636.
\newblock \href {http://dx.doi.org/10.1063/1.1672048}
  {\path{doi:10.1063/1.1672048}}.

\bibitem{boublik1974hard}
T.~Boubl{\'i}k, Hard convex body equation of state, J. Chem. Phys. 63~(9)
  (1975) 4084.
\newblock \href {http://dx.doi.org/10.1063/1.431882}
  {\path{doi:10.1063/1.431882}}.

\bibitem{frenkel1997}
P.~Bolhuis, D.~Frenkel, Tracing the phase boundaries of hard spherocylinders,
  J. Chem. Phys. 106~(2) (1997) 666--687.
\newblock \href {http://dx.doi.org/10.1063/1.473404}
  {\path{doi:10.1063/1.473404}}.

\bibitem{holovko1991effects}
M.~F. Holovko, Y.~V. Kalyuzhnyi, On the effects of association in the
  statistical theory of ionic systems. {A}nalytic solution of the {PY}-{MSA}
  version of the {W}ertheim theory, Mol. Phys. 73~(5) (1991) 1145--1157.
\newblock \href {http://dx.doi.org/10.1080/00268979100101831}
  {\path{doi:10.1080/00268979100101831}}.

\bibitem{Blum95}
L.~Blum, O.~Bernard, The general solution of the binding mean spherical
  approximation for pairing ions, J. Stat. Phys. 79~(3--4) (1995) 569--583.
\newblock \href {http://dx.doi.org/10.1007/bf02184871}
  {\path{doi:10.1007/bf02184871}}.

\bibitem{Protsykevytch1997}
I.~Protsykevytch, Y.~Kalyuzhnyi, M.~Holovko, L.~Blum, Solution of the polymer
  mean spherical approximation for the totally flexible sticky two-point
  electrolyte model, J. Mol. Liq. 73-74 (1997) 1--20.
\newblock \href {http://dx.doi.org/10.1016/s0167-7322(97)00053-6}
  {\path{doi:10.1016/s0167-7322(97)00053-6}}.

\bibitem{KALYUZHNYI1998}
Y.~Kalyuzhnyi, Thermodynamics of the polymer mean-spherical ideal chain
  approximation for a fluid of linear chain molecules, Mol. Phys. 94~(4) (1998)
  735--742.
\newblock \href {http://dx.doi.org/10.1080/00268979809482366}
  {\path{doi:10.1080/00268979809482366}}.

\bibitem{Bernard96}
O.~Bernard, L.~Blum, Binding mean spherical approximation for pairing ions:
  {A}n exponential approximation and thermodynamics, J. Chem. Phys. 104~(12)
  (1996) 4746--4754.
\newblock \href {http://dx.doi.org/10.1063/1.471168}
  {\path{doi:10.1063/1.471168}}.

\bibitem{Jiang02}
J.~Jiang, L.~Blum, O.~Bernard, J.~M. Prausnitz, S.~I. Sandler, Criticality and
  phase behavior in the restricted-primitive model electrolyte: {D}escription
  of ion association, J. Chem. Phys. 116~(18) (2002) 7977--7982.
\newblock \href {http://dx.doi.org/10.1063/1.1468638}
  {\path{doi:10.1063/1.1468638}}.

\bibitem{Ebeling1968}
W.~Ebeling, Zur theorie der bjerrumschen ionenassoziation in elektrolyten,
  Zeitschrift f\"{u}r Physikalische Chemie 238O~(1) (1968) 400–402.
\newblock \href {http://dx.doi.org/10.1515/zpch-1968-23847}
  {\path{doi:10.1515/zpch-1968-23847}}.

\bibitem{Olaussen91}
K.~Olaussen, G.~Stell, New microscopic approach to the statistical mechanics of
  chemical association, J. Stat. Phys. 62~(1--2) (1991) 221--237.
\newblock \href {http://dx.doi.org/10.1007/bf01020867}
  {\path{doi:10.1007/bf01020867}}.

\bibitem{raineri2000phase}
F.~Raineri, J.~Routh, G.~Stell, Phase separation in the size-asymmetric
  primitive model, Le Journal de Physique IV 10~(PR5) (2000) Pr5--99.
\newblock \href {http://dx.doi.org/10.1051/jp4:2000511}
  {\path{doi:10.1051/jp4:2000511}}.

\bibitem{bernard1996binding}
O.~Bernard, L.~Blum, Binding mean spherical approximation for pairing ions: an
  exponential approximation and thermodynamics, J. Chem. Phys. 104~(12) (1996)
  4746--4754.
\newblock \href {http://dx.doi.org/10.1063/1.471168}
  {\path{doi:10.1063/1.471168}}.

\bibitem{Bernard2000}
O.~Bernard, L.~Blum, Thermodynamics of a model for flexible polyelectrolytes in
  the binding mean spherical approximation, J. Chem. Phys. 112~(16) (2000)
  7227--7237.
\newblock \href {http://dx.doi.org/10.1063/1.481287}
  {\path{doi:10.1063/1.481287}}.

\bibitem{Kalyuzhnyi2001}
Y.~V. Kalyuzhnyi, P.~T. Cummings, Multicomponent mixture of charged hard-sphere
  chain molecules in the polymer mean-spherical approximation, J. Chem. Phys.
  115~(1) (2001) 540--551.
\newblock \href {http://dx.doi.org/10.1063/1.1376426}
  {\path{doi:10.1063/1.1376426}}.

\bibitem{Zhou_SPM}
Y.~Zhou, G.~Stell, Criticality of charged systems. {II}. the binary mixture of
  hard spheres and ions, J. Chem. Phys. 102~(14) (1995) 5796--5802.
\newblock \href {http://dx.doi.org/10.1063/1.469311}
  {\path{doi:10.1063/1.469311}}.

\bibitem{Kalyuzhnyi1998:2}
Y.~V. Kalyuzhnyi, C.-T. Lin, G.~Stell, Primitive models of chemical
  association. iv. polymer percus–yevick ideal-chain approximation for
  heteronuclear hard-sphere chain fluids, J. Chem. Phys. 108~(15) (1998)
  6525–6534.
\newblock \href {http://dx.doi.org/10.1063/1.476059}
  {\path{doi:10.1063/1.476059}}.

\bibitem{Rebelo2005}
L.~P.~N. Rebelo, J.~N. Canongia~Lopes, J.~M. S.~S. Esperan\c{c}a, E.~Filipe, On
  the critical temperature, normal boiling point, and vapor pressure of ionic
  liquids, J. Phys. Chem. B 109~(13) (2005) 6040–6043.
\newblock \href {http://dx.doi.org/10.1021/jp050430h}
  {\path{doi:10.1021/jp050430h}}.

\bibitem{camp1999ion}
P.~J. Camp, G.~Patey, Ion association and condensation in primitive models of
  electrolyte solutions, The Journal of chemical physics 111~(19) (1999)
  9000--9008.

\bibitem{Pizio04}
O.~Pizio, A.~Patrykiejew, S.~Sokolowski, Towards the description of the phase
  behaviour of electrolyte solutions in slit-like pores. {D}ensity functional
  approach for the restricted primitive model, Condens. Matter Phys. 7~(4)
  (2004) 779--792.
\newblock \href {http://dx.doi.org/10.5488/cmp.7.4.779}
  {\path{doi:10.5488/cmp.7.4.779}}.

\bibitem{parsons1979nematic}
J.~D. Parsons, Nematic ordering in a system of rods, Phys. Rev. A 19~(3) (1979)
  1225--1230.
\newblock \href {http://dx.doi.org/10.1103/physreva.19.1225}
  {\path{doi:10.1103/physreva.19.1225}}.

\bibitem{lee1987numerical}
S.-D. Lee, A numerical investigation of nematic ordering based on a simple
  hard-rod model, J. Chem. Phys. 87~(8) (1987) 4972--4974.
\newblock \href {http://dx.doi.org/10.1063/1.452811}
  {\path{doi:10.1063/1.452811}}.

\end{thebibliography}





\end{document}